\documentclass[aps,preprint,floatfix,showpacs]{revtex4}
\usepackage{graphicx}
\usepackage{amssymb,amsmath,bbm,mathrsfs}
\usepackage{t1enc}
\input{epsf}
\begin{document}
\newcommand{\ket}[1]{\vert #1 \rangle}
\newcommand{\bra}[1]{\langle #1 \vert}
\newcommand{\dket}[1]{\vert #1 \rangle\rangle}
\newcommand{\dbra}[1]{\langle\langle #1 \vert}
\newcommand{\braket}[2]{\langle #1 \vert #2 \rangle}
\newcommand{\dbraket}[2]{\langle\langle #1 \vert #2 \rangle\rangle}
\newcommand{\ketbra}[2]{\vert #1 \rangle \! \langle #2 \vert}
\newcommand{\media}[1]{\langle #1 \rangle}
\newcommand{\Tr}{{\rm Tr}}
\newcommand{\calpha}{\alpha^*}
\newcommand{\cbeta}{\beta^*}
\newcommand{\cxi}{\xi^*}
\newcommand{\czeta}{\zeta^*}
\newcommand{\cv}{v^*}
\newcommand{\cz}{z^*}
\newcommand{\cw}{w^*}
\newcommand{\iid}{\mathbb{I}}
\newcommand{\wtA}{A}
\newcommand{\wtB}{B}
\newcommand{\bmsigma}{\boldsymbol \sigma}
\newcommand{\bmX}{\boldsymbol X}
\newcommand{\bbGamma}{{\rm I}\!\Gamma}
\newcommand{\calN}{{\cal N}}
\newcommand{\Bop}{{\mathscr B}}
\newcommand{\Oop}{{\mathscr O}}
\newcommand{\Eop}{{\mathscr E}}
\title{Effect of noise and enhancement of nonlocality in on/off photodetection}
\author{Carmen Invernizzi}
\author{Stefano Olivares}
\email{Stefano.Olivares@mi.infn.it}
\author{Matteo G. A. Paris}
\affiliation{Dipartimento di
Fisica dell'Universit\`a degli Studi, Milano, Italia.}
\author{Konrad Banaszek}
\affiliation{Institute of Physics, Nicolaus Copernicus University,
ul.~Grudzi\k{a}dzka 5, PL-87-100 Toru\'{n}, Poland}
\begin{abstract}
Nonlocality of two-mode states of  light is addressed by means 
of CHSH inequality based on displaced on/off photodetection. Effects 
due to non-unit quantum efficiency and nonzero dark counts are taken 
into account. Nonlocality of both balanced and unbalanced superpositions 
of few photon-number states, as well as that of multiphoton twin beams, 
is investigated. We find that unbalanced superpositions show larger 
nonlocality than balanced one when noise affects the photodetection 
process. De-Gaussification by means of (inconclusive) photon subtraction is 
shown to enhance nonlocality of twin beams in the low energy regime. 
We also show that when the measurement is described by a POVM, rather than 
a set of projectors, the maximum achievable value of the Bell parameter in 
the CHSH inequality is decreased, and is no longer given by the Cirel'son 
bound. 
\end{abstract}
 \maketitle
\section{Introduction}
Quantum entanglement for both discrete and continuous variable systems 
has been extensively analyzed, also revealing its subtle relations with 
other quantum mechanical features such as nonlocality. Indeed, it has
been pointed out that the concept of entanglement coincides
with nonlocality only for the simple case of bipartite pure states. As soon
as we deal with mixed states, entangled states can be found which do
not show properties of nonlocality while, not unexpectedly, the converse 
is always true \cite{wer89}.
In addition, the amount of nonlocality, i.e., the amount of
violation of a suitable Bell inequality, crucially depends on the 
nonlocality test adopted in the analysis, ranging from no violation to
maximal violation for the same (entangled) quantum state.
\par
In this paper we address nonlocality of different kinds of two-mode
states of light by means of {\em displaced} on/off photodetection taking
into account the effects of non-unit quantum efficiency and dark counts.
This kind of measurement was first proposed in Ref.~\cite{BW:PRL:99},
where, in particular, it was pointed out that the correlation functions
violating the Bell inequalities (in the ideal case) involve the joint 
two-mode $Q$-function.
The reason to pay a particular attention to on/off tests of nonlocality
is twofold. On one hand, it has been shown that violation of Bell
inequalities may be quite pronounced for some relevant state of light 
in the ideal case \cite{BW:PRL:99}. On the other
hand, and more importantly, on/off tests may be effectively implemented
with currently technology. In this framework, it is of interest 
to take into account the effects of experimental imperfections, e.g., 
non-unit quantum efficiency and nonzero dark counts \cite{bana:pra:02}, and to 
investigate the nonlocality properties of physically realizable entangled 
states. Indeed, realistic  implementations of quantum information protocols
require the investigation of nonlocality properties of quantum 
states in a noisy environment. In particular, the robustness of 
nonlocality should be addressed, as well as the design of protocols 
to preserve and possibly enhance nonlocality in the presence of noise.
\par
The paper is structured as follows: in the next Section we describe
in some detail the nonlocality test we use throughout the paper, 
while in Section \ref{s:bbasis} we analyze nonlocality of superpositions, 
both balanced  (Bell states) and unbalanced, involving zero- and
one-photon states. In Section \ref{s:twop} we address
the nonlocality of superpositions containing two-photon states, 
whereas Section \ref{s:TWB} is focused on multiphoton twin-beam 
state. In Section \ref{s:DarkC} we analyze the effect of dark counts
on the violation of CHSH inequality, whereas in Section \ref{s:IPS}
we address inconclusive photon subtraction (IPS) as a method to enhance
nonlocality of twin-beam. 
Section \ref{s:choice} is devoted to a more detailed analysis of the 
choice of parametrization leading to violation of inequalities, 
whereas, in Section \ref{s:cirel} we show how a nonlocality test based on
POVM measurement cannot yield the maximal violation of inequalities 
expressed by the Cirel'son bound. Finally, Section \ref{s:remarks}
closes the paper with some concluding remarks. 
\section{Correlation functions and Bell parameter}
\begin{figure}[h]
\includegraphics[width=85mm]{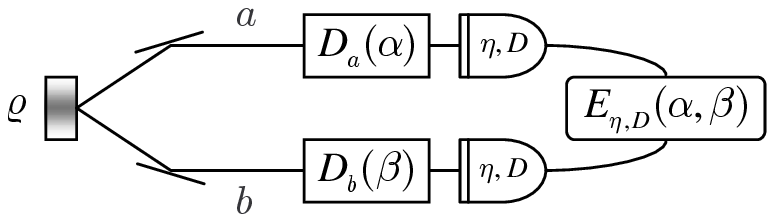}
\vspace{-.5cm}
\caption{Scheme of the nonlocality test based on displaced 
on/off photodetection: the two modes $a$ and $b$ of a bipartite state 
$\varrho$ are locally displaced by an amount $\alpha$ and $\beta$ 
respectively, and then revealed through on/off photodetection.
The corresponding correlation function violates Bell's
inequalities for dichotomous measurements for a suitable choice of the 
parameters $\alpha$ and $\beta$, depending on the kind of state
under investigation. The violation holds also for non-unit quantum 
efficiency and non-zero dark counts.
}\label{f:Donoff}
\end{figure}

The nonlocality test we are going to analyze is schematically depicted in
Fig.~\ref{f:Donoff}: two modes of the radiation field, $a$ and $b$, are
excited in a given  (entangled) two-mode state described by the density
matrix $\varrho$, and then are locally displaced by an amount $\alpha$ and
$\beta$ respectively. Finally, the two modes are revealed by on/off
photodetectors, i.e., detectors which have no output when no photon is
detected and a fixed output when one or more photons are detected. The
action of an on/off detector is described by the following two-value
positive operator-valued measure (POVM) $\{\Pi_{0,\eta,D},
\Pi_{1,\eta,D}\}$ \cite{FOP:2005}
\begin{subequations}
\label{povm1}
\begin{align}
{\Pi}_{0,\eta,D} &= \frac1{1+D}\: \sum_{k=0}^{\infty}
\left( 1-\frac{\eta}{1+D} \right)^k \ket{k}\bra{k}\:,\\ 
{\Pi}_{1,\eta,D} &= \mathbb{I} - {\Pi}_{0,\eta,D}\:,
\end{align}
\end{subequations}
$\eta$ being the quantum efficiency and $D$ the mean number 
of dark counts, i.e., of clicks with vacuum input. 
In writing Eq. (\ref{povm1}) we have considered a thermal background
as the origin of dark counts. An analogous expression may be written
for a Poissonian background (see Appendix \ref{a:onoff}). For small 
values of the mean number $D$ of dark counts (as it generally happens at 
optical frequencies) the two kinds of background are indistinguishable.
\par
Overall, taking into account the displacement, the measurement 
on both modes $a$ and $b$ is described by the POVM (we are 
assuming the same quantum efficiency and dark counts for both the
photodetectors)
\begin{equation}
{\Pi}_{hk}^{(\eta,D)} (\alpha,\beta)=
{\Pi}_{h}^{(\eta,D)}(\alpha)\,
\otimes
{\Pi}_{k}^{(\eta,D)}(\beta)\,
\label{povmhk}\;,
\end{equation}
where $h,k = 0,1$, and ${\Pi}_{h}^{(\eta,D)}(z)\equiv  D(z)\,
{\Pi}_{h,\eta,D}\,D^{\dag}(z)$, $D(z)=\exp\left\{z a^\dag - z^* a\right\}$ 
being the displacement operator and $z\in {\mathbb C}$ a complex
parameter.
\par
In order to analyze the nonlocality of the state $\varrho$,
we introduce the following correlation function:
\begin{align}
E_{\eta,D}(\alpha,\beta) &=
\sum_{h,k=0}^{1} (-)^{h+k}\, 
\left\langle {\Pi}_{hk}^{(\eta,D)} (\alpha,\beta) 
\right\rangle \label{E:eta} \\
&= 1 + 4\, {\cal I}_{\eta,D}(\alpha,\beta) - 
2\, \left[ {\cal G}_{\eta,D}(\alpha) + {\cal Y}_{\eta,D}(\beta) \right]\:,
\nonumber
\end{align}
where
\begin{subequations}
\label{DefFuns}
\begin{align}
&{\cal I}_{\eta,D}(\alpha,\beta) = 
\left\langle {\Pi}_{00}^{(\eta,D)} (\alpha,\beta) \right\rangle  \\
&{\cal G}_{\eta,D}(\alpha) = 
\left\langle {\Pi}_{0}^{(\eta,D)}(\alpha)\otimes\mathbb{I} \right\rangle 
\\ 
&{\cal Y}_{\eta,D}(\beta) = 
\left\langle \mathbb{I}\otimes{\Pi}_{0}^{(\eta,D)}(\beta) \right\rangle\:,
\end{align}
\end{subequations}
and where $\media{ A } \equiv {\rm Tr}[\varrho\, A]$ denotes
ensemble average on both the modes.
The so-called  Bell parameter is defined by considering four different
values of the complex displacement parameters as follows
\begin{align}
{\cal B}_{\eta,D} =&\: \: E_{\eta,D}(\alpha,\beta) +
E_{\eta,D}(\alpha',\beta) \nonumber + E_{\eta,D}(\alpha,\beta')  \\ 
&- E_{\eta,D}(\alpha',\beta')\\
=& \:\: 2 + 4\left\{ {\cal I}_{\eta,D}(\alpha,\beta) 
+ {\cal I}_{\eta,D}(\alpha',\beta) +
{\cal I}_{\eta,D}(\alpha,\beta')\right. \nonumber\\
&\left.
- {\cal I}_{\eta,D}(\alpha',\beta')
- {\cal G}_{\eta,D}(\alpha) - {\cal Y}_{\eta,D}(\beta) \right\}\,.
\label{B:param}
\end{align}
Any local theory implies that $|{\cal B}_{\eta,D}|$ satisfies the 
CHSH version of the Bell inequality, i.e., $|{\cal B}_{\eta,D}|\leq 2$ 
$\forall \alpha,\alpha',\beta, \beta'$ \cite{CHSH}, while
quantum mechanical description of the same kind of experiments does not
impose this bound (see Section \ref{s:cirel} for more details on
quantum-mechanical bounds on $|{\cal B}_{\eta,D}|$ in on/off
experiments).
\par
Notice that using Eqs.~(\ref{povm1}) and (\ref{DefFuns}) we obtain the 
following scaling properties for the functions ${\cal
I}_{\eta,D}(\alpha,\beta)$, ${\cal G}_{\eta,D}(\alpha)$ and 
${\cal Y}_{\eta,D}(\beta)$
\begin{subequations}
\label{D2eta}
\begin{align}
&{\cal I}_{\eta,D}(\alpha,\beta) = \left(\frac{1}{1+D}\right)^2\,
{\cal I}_{\eta/(1+D)}(\alpha,\beta)\\
&{\cal G}_{\eta,D}(\alpha) = \frac{1}{1+D}\,{\cal G}_{\eta/(1+D)}(\alpha)\\
&{\cal Y}_{\eta,D}(\beta) = \frac{1}{1+D}\,{\cal Y}_{\eta/(1+D)}(\beta)
\end{align}
\end{subequations}
where ${\cal I}_{\eta} = {\cal I}_{\eta,0}$,
${\cal G}_{\eta} = {\cal G}_{\eta,0}$, and
${\cal Y}_{\eta} = {\cal Y}_{\eta,0}$.
Therefore, it will be enough to study the Bell parameter
for $D=0$, namely ${\cal B}_{\eta} = {\cal B}_{\eta,0}$, and then we can use
Eqs.~(\ref{D2eta}) to take into account the effects of non negligible
dark counts. From now on we will assume $D=0$ and suppress the
explicit dependence on $D$. Notice that using expression (\ref{B:param}) for
the Bell parameter the CHSH inequality $|{\cal B}_{\eta,D}|\leq 2$ can be
rewritten as 
\begin{align}
-1 <& \:\: 
{\cal I}_{\eta,D}(\alpha,\beta) + 
{\cal I}_{\eta,D}(\alpha',\beta) +
{\cal I}_{\eta,D}(\alpha,\beta') \nonumber \\ &-
{\cal I}_{\eta,D}(\alpha',\beta') 
- {\cal G}_{\eta,D}(\alpha) - {\cal Y}_{\eta,D}(\beta) < 0
\label{CH}\;,
\end{align}
which represents the CH version of the Bell inequality for our system \cite{CH}.
\par
In order to simplify the calculations, throughout this paper we will use the
Wigner formalism. The Wigner functions associated with the
elements of the POVM (\ref{povm1}) for $D=0$ are given by 
(see Appendix \ref{a:onoff})
\begin{eqnarray}
W[\Pi_{0,\eta}] (z) &=& \frac{\Delta_\eta}{\pi \eta}\,
\exp\left\{ - \Delta_\eta\, |z|^2  \right\}\,, \\
W[\Pi_{1,\eta}] (z) &=& W[\iid](z) - W[\Pi_{0,\eta}] (z) \,,
\end{eqnarray}
with $\Delta_\eta = 2 \eta / (2 - \eta)$, and $W[\iid](z)=\pi^{-1}$.
Then, noticing that for any operator $O$ one has 
\begin{equation}
W[D(\alpha)\,O \,D^{\dag}(\alpha)](z) = W[O](z - \alpha)\,,
\end{equation}
it follows that
$W[D(\alpha)\,{\Pi}_{0,\eta}\,D^{\dag}(\alpha)](z)$ is given by
\begin{equation}
W[D(\alpha)\,{\Pi}_{0,\eta}\,D^{\dag}(\alpha)](z)
= W[\Pi_{0,\eta}] (z-\alpha)\,,
\end{equation}
and therefore
\begin{align}
W[\Pi_{00}^{(\eta,0)}(\alpha,\beta)](z,w)&= 
W[\Pi_{0,\eta}] (z-\alpha)\nonumber\\
&\hspace{1cm}\times W[\Pi_{0,\eta}] (w-\beta) \\
W[\Pi_{0,\eta}(\alpha)\otimes\iid](z,w)&= 
W[\Pi_{0,\eta}] (z-\alpha)\: \pi^{-1} \\
W[\iid \otimes \Pi_{0,\eta}(\beta)](z,w)&= 
 \pi^{-1}\:W[\Pi_{0,\eta}] (w-\beta) 
\label{uwigs}\;.
\end{align}
Finally, thanks to the trace rule expressed in the phase space
of two modes, i.e.,
\begin{equation}
{\rm Tr}[O_1\,O_2] =
\pi^2 
\int_{\mathbb{C}^2}\!d^{2}z \:d^{2}w
\, W[O_1](z,w)\,W[O_2](z,w)\,,
\end{equation}
one can evaluate the functions ${\cal I}_{\eta}(\alpha,\beta)$, 
${\cal G}_{\eta}(\alpha)$, and ${\cal Y}_{\eta}(\beta)$, and in turn 
the Bell parameter ${\cal B}_{\eta}$ in Eq.~(\ref{B:param}),
as a sum of Gaussian integrals in the complex plane.
\section{Nonlocality of the Bell states}\label{s:bbasis}
We start our analysis by considering balanced superpositions of
of zero- and one-photon states, i.e., the so-called 
Bell states, which are described by the density matrices
\begin{equation}
\varrho_{\pm} = \ket{\Psi_{\pm}}\bra{\Psi_{\pm}}\,,\quad
\sigma_{\pm} = \ket{\Phi_{\pm}}\bra{\Phi_{\pm}}\,
\end{equation}
where
\begin{align}
\ket{\Psi_{\pm}} &= \frac{1}{\sqrt{2}}
\big(\ket{1}\ket{0}\pm\ket{0}\ket{1}\big)\,,\\
\ket{\Phi_{\pm}} &= \frac{1}{\sqrt{2}}
\big(\ket{0}\ket{0}\pm\ket{1}\ket{1}\big)\,.
\end{align}
In optical implementations Bell states 
$\ket{\Psi_{\pm}}$ are obtained from single-photon sources using
linear optical elements, while preparation of
$\ket{\Phi_{\pm}}$ requires active devices based on spontaneous
parametric down-conversion. \par 
The Wigner functions of the Bell states are given by 
\begin{align}
W[\varrho_{\pm}](z,w) =& \:\:
\frac{4}{\pi^2}\,\exp\left\{-2|z|^2-2|w|^2\right\} \nonumber \\
&\times \left( 2\left|z \mp w \right|^2 -1 \right)\,,
\end{align}
and
\begin{align}
W[\sigma_{\pm}](z,w) = & \:\:
\frac{4}{\pi^2}\,\exp\left\{-2|z|^2-2|w|^2\right\} \nonumber \\
&\times 
\left(1 - 2\left|z\mp w\right|^2 + 8\,|z|^2 |w|^2\right)\,,
\end{align}
respectively.
\par
\begin{figure}[h]
\vspace{4cm}
\setlength{\unitlength}{.4cm} 
\centerline{%
\begin{picture}(0,0) 
\put(0,0){\makebox(0,0)[c]{\epsfxsize=15\unitlength\epsffile{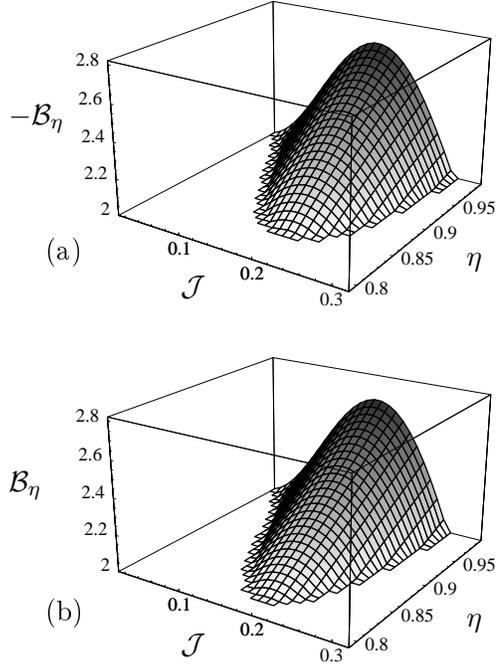}}}
\put(-9.2,7){$-{\cal B}_{\eta}$}
\put(-3.5,1.3){${\cal J}$}
\put(6,2.3){$\eta$}
\put(-9.2,-5.2){${\cal B}_{\eta}$}
\put(-3.5,-10.6){${\cal J}$}
\put(6,-9.7){$\eta$}
\put(-8,2.5){(a)}
\put(-8,-9.5){(b)}
\end{picture} }
\vspace{4cm}
\caption{ Plot of $-{\cal B}_{\eta}$ for the states $\ket{\Psi_\pm}$ (a)
and of ${\cal B}_{\eta}$ for $\ket{\Phi_\pm}$ (b) as functions of ${\cal
J}$ and $\eta$. The maximum violations  for $\eta=1$ are: (a) $-{\cal
B}_{\eta}=2.69$, and (b) ${\cal B}_{\eta}=2.68$, which are both obtained
when ${\cal J}=0.17$.  For the particular choice of the parametrizations
${\cal B}_{\eta}$ is the same for $\ket{\Psi_\pm}$ and for
$\ket{\Phi_\pm}$.} \label{f:PsiPhi}
\end{figure}

Let us first consider $\varrho_{\pm}$. In this case the 
functions in Eqs. (\ref{DefFuns}) are given by 
\begin{align}
&{\cal I}_{\eta}(\alpha,\beta) = \frac12\,
e^{-\eta(|\alpha|^2-|\beta|^2)} \left[ 2(1-\eta) + 
\eta^2 |\alpha\mp \beta|^2 \right]\\
&{\cal G}_{\eta}(\alpha) = {\cal Y}_{\eta}(\alpha) = \frac12\,
e^{-\eta|\alpha|^2}\left[2-\eta+\eta^2 |\alpha|^2 \right]\,,\label{G:Y:Psi}
\end{align}
while the Bell's parameter is obtained using Eq. (\ref{B:param}). 
Maximization of $|B_\eta|$, carried out using both analytical and 
numerical methods, indicates that the imaginary parts of  the 
parameters $\alpha,\alpha',\beta,\beta'$ can be neglected 
for $\eta=1$, while it influences only slightly the value 
of $|B_\eta|$ for $\eta <1$.  More details about the choice of the 
parametrization are given in Sec.~\ref{s:choice}.
Using the parameterizations: $\alpha = -\beta
= {\cal J}$, $\alpha' = -\beta' = -\sqrt{11}{\cal J}$ for the state
$\ket{\Psi_{+}}$, and
$\alpha = \beta = {\cal J}$, $\alpha' = \beta'
= -\sqrt{11}{\cal J}$ for the state $\ket{\Psi_{-}}$ with 
${\cal J} \in {\mathbb R}$ we get 
the same Bell's parameter for both the states and a maximum
violation $|B|= 2.68$ (when $\eta = 1$).
The Bell's parameter for $\varrho_\pm$ is shown in 
Fig.~\ref{f:PsiPhi} (a) as a function of ${\cal J}$ and $\eta$.
\par
If we consider $\sigma_{\pm}$, we have
\begin{align}
{\cal I}_{\eta}(\alpha,\beta) = & \frac12\,
e^{-\eta(|\alpha|^2+|\beta|^2)}\,\big\{ 2(1-\eta)
+ \eta^2 \big[ 1 + |\alpha\pm \beta^*|^2 \nonumber\\ 
& + (1-\eta|\alpha|^2)(1-\eta|\beta|^2) \big]\big\}\,,
\end{align}
whereas ${\cal G}_{\eta}(\alpha)$ and ${\cal Y}_{\eta}(\beta)$
are given in
Eq.~(\ref{G:Y:Psi}). As for the states $\ket{\Psi_{\pm}}$, 
the optimal parametrization has been obtained by a semi-analytical 
analysis. We get $\alpha = -\beta =
{\cal J}$, $\alpha' = -\beta' = -\sqrt{11}{\cal J}$ for the state
$\ket{\Phi_{+}}$, and $\alpha = \beta = {\cal J}$, $\alpha' = \beta' =
-\sqrt{11}{\cal J}$ for the state $\ket{\Phi_{-}}$ (see Sec.~\ref{s:choice}
for more details).  Thanks to this choice, ${\cal B}_{\eta}$ is maximized
(when $\eta = 1$) for both the Bell states $\sigma_\pm$.
The results are shown in Fig.~\ref{f:PsiPhi} (b).
\par
The overall effect of non-unit quantum efficiency is to reduce the
interval of ${\cal J}$ values in which there is violation. Notice 
that the states $\ket{\Phi_{\pm}}$ are slightly more robust than 
the $\ket{\Psi_{\pm}}$ one. In fact, one has $|{\cal B}_{\eta}|\le 2$.
as far as $\eta$ falls below $83.6\%$ for $\ket{\Psi_{\pm}}$ and 
$81.6\%$ for $\ket{\Phi_{\pm}}$. These results are consistent with the 
study given in Ref.~\cite{bana:pra:02}, where the authors also have taken 
into account mode mismatch and have used a numerical algorithm in order 
to find the best choice of the parameters $\alpha$, $\beta$, $\alpha'$, and
$\beta'$.
\subsection{Unbalanced superpositions}
\begin{figure}[h]
\vspace{4cm}
\setlength{\unitlength}{.5cm} 
\centerline{%
\begin{picture}(0,0) 
\put(0,0){\makebox(0,0)[c]{\epsfxsize=12\unitlength\epsffile{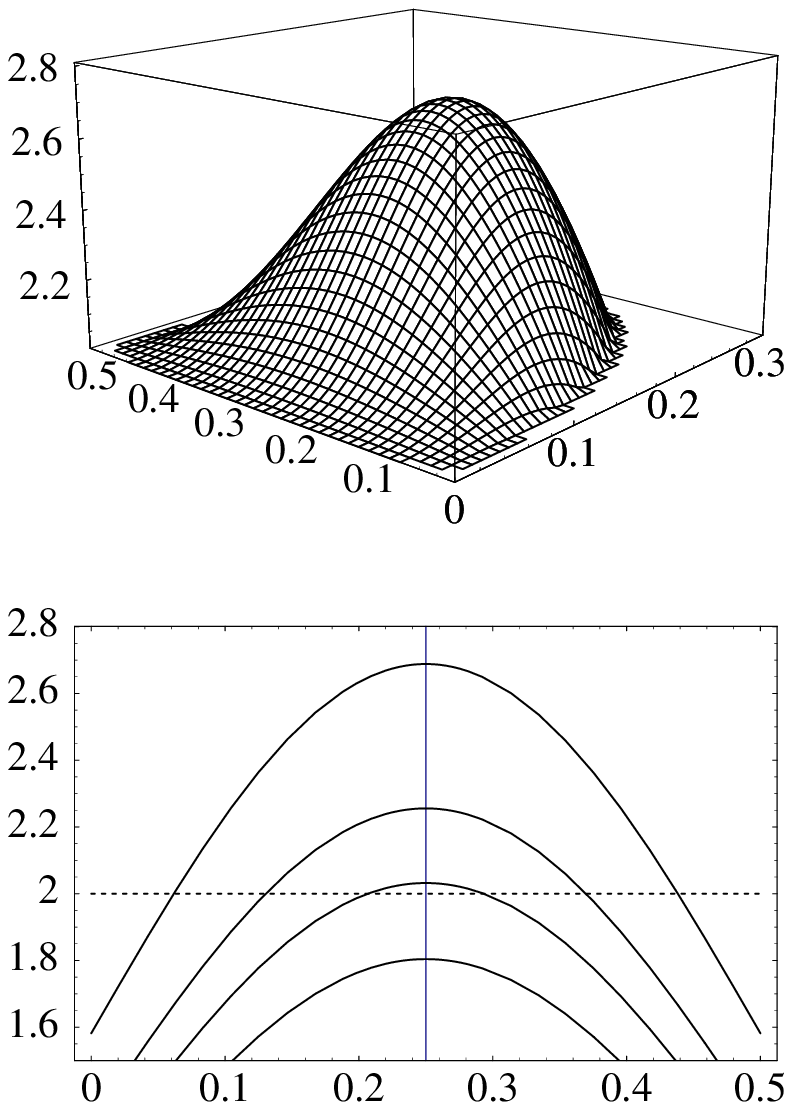}}}
\put(-7.4,5){$-{\cal B}_{\eta}$}
\put(4,1){${\cal J}$}
\put(-3.5,1){$\varphi/\pi$}
\put(-7.4,-4){$-{\cal B}_{\eta}$}
\put(0,-8.6){$\varphi/\pi$}
\put(-6,2){(a)}
\put(-6,-8.5){(b)}
\end{picture}
}
\vspace{4.2cm}
\caption{Plot of $-{\cal B}_{\eta}$: (a) for $\ket{\Psi_{\varphi}}$
as a function of ${\cal J}$ and $\varphi$ in the case of ideal (i.e.,
$\eta = 1$) on/off photodetection; (b) for ${\cal J} = 0.17$ and
different values of $\eta$: from bottom to top $\eta = 1.0$,
$0.9$, $0.85$, and $0.8$. The vertical line is $\varphi = \pi/4$.}
\label{f:PhiGen}
\end{figure}

\begin{figure}[h]
\vspace{4cm}
\setlength{\unitlength}{.5cm} 
\centerline{%
\begin{picture}(0,0) 
\put(.3,.3){\makebox(0,0)[c]{\epsfxsize=12\unitlength\epsffile{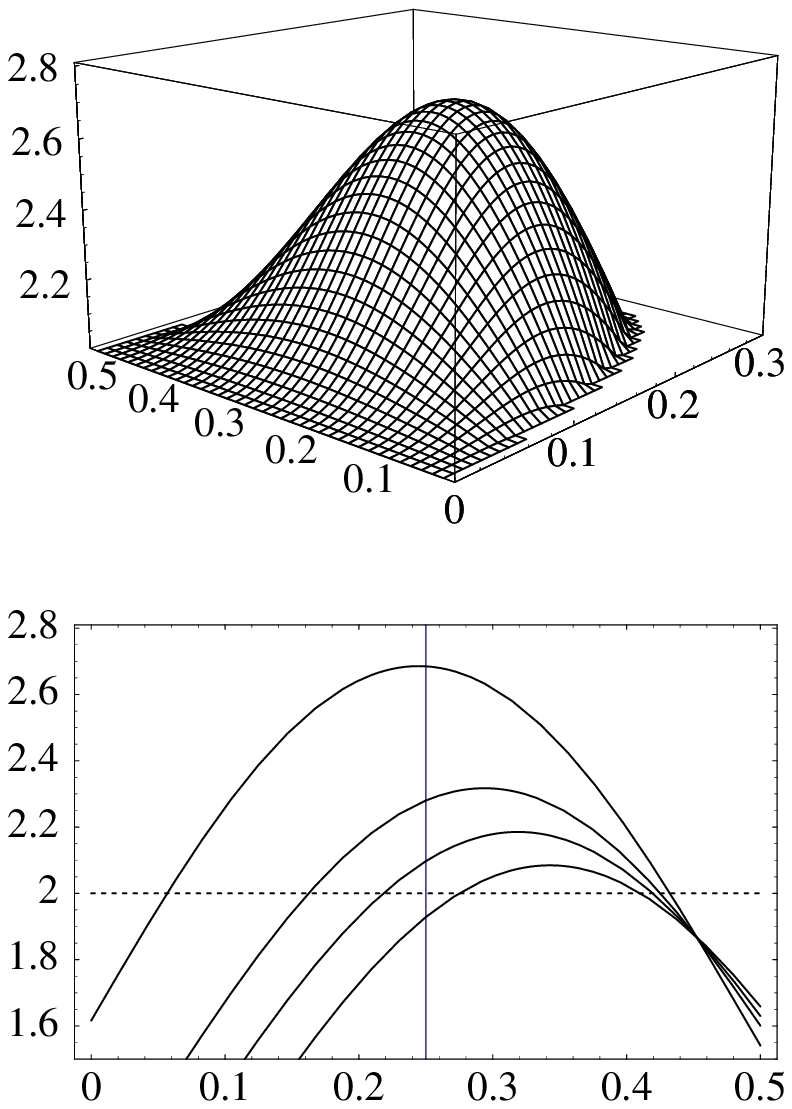}}}
\put(-6.5,5){${\cal B}_{\eta}$}
\put(4,1.2){${\cal J}$}
\put(-3.5,1.2){$\varphi/\pi$}
\put(-6.5,-4){${\cal B}_{\eta}$}
\put(0,-8.2){$\varphi/\pi$}
\put(-6,2){(a)}
\put(-6,-8){(b)}
\end{picture}
}
\vspace{4cm}
\caption{Plot of ${\cal B}_{\eta}$: (a) for $\ket{\Phi_{\varphi}}$
as a function of ${\cal J}$ and $\varphi$ in the case of ideal (i.e.,
$\eta = 1$) on/off photodetection; (b) for ${\cal J} = 0.17$ and
different values of $\eta$: from top to bottom $\eta = 1.0$,
$0.9$, $0.85$, $0.8$, and $0.75$. The vertical line is $\varphi = \pi/4$.}
\label{f:PsiGen}
\end{figure}

Our analysis of on-off photodetection is aimed to describe optical 
implementations of nonlocality tests, where most of the experiments 
have been realized. In this framework the Bell states 
$\ket{\Psi_{\pm}}$ may be obtained from single-photon 
sources using balanced beam-splitters. In order to take into 
account possible imperfections it is worth to analyze nonlocality
properties of the class of states that can be obtained from unbalanced
beam splitters. Indeed, the analysis given above can be extended in order 
to describe  general superpositions of the form
\begin{align}
\ket{\Psi_{\varphi}} &= 
\sin\varphi \ket{1}\ket{0} + \cos\varphi \ket{0}\ket{1}\,,\\
\ket{\Phi_{\varphi}} &= 
\sin\varphi \ket{0}\ket{0} + \cos\varphi \ket{1}\ket{1}\,.
\end{align}
Since the calculations are similar to the ones of the Bell states, here we
do not explicitly write the analytical results for the states
$\ket{\Psi_{\varphi}}$ and $\ket{\Phi_{\varphi}}$. Rather, we plot the
corresponding Bell parameter ${\cal B}_{\eta}$ in Figs.~\ref{f:PhiGen} and
\ref{f:PsiGen}. In both the plots we used the same parametrization as for
the Bell states. As one can see in Fig.~\ref{f:PhiGen}, in the case of the
superposition $\ket{\Psi_{\varphi}}$ the best result are obtained for the
balanced superposition, namely $\varphi = \pi / 4$. On the other hand, 
the case of $\ket{\Phi_{\varphi}}$ shows a different behavior: here the 
maximum of the violation for the ideal
case ({\em i. e. } $\eta = 1$) is achieved for a value of $\varphi$
slightly smaller than $\pi / 4$, and it increases as the detection
efficiency decreases. Moreover, by the comparison between
Fig.~\ref{f:PsiPhi} (b) and Fig.~\ref{f:PhiGen}, we can see that, for the
particular choice of the parametrization, when $\eta
= 0.8$ the balanced superposition does not violates the CHSH inequality,
whereas, adjusting the parameter $\varphi$, the unbalanced superposition
violates it and it does until the efficiency falls below the threshold value
$\eta \approx 0.74$.
\section{Nonlocality of superpositions containing two-photon states}
\label{s:twop}
\begin{figure}[h]
\vspace{2.5cm}
\setlength{\unitlength}{.5cm} 
\centerline{%
\begin{picture}(0,0) 
\put(0,0){\makebox(0,0)[c]{\epsfxsize=12\unitlength\epsffile{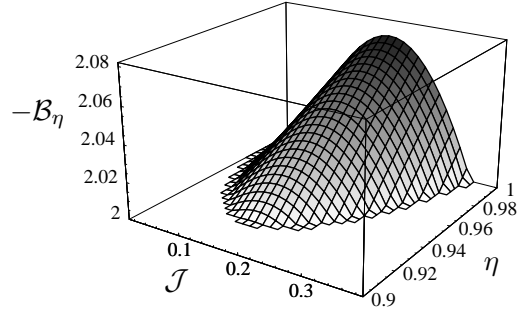}}}
\put(-7.6,1){$-{\cal B}_{\eta}$}
\put(5,-3){$\eta$}
\put(-3.5,-3.5){${\cal J}$}
\end{picture}
}
\vspace{2cm}
\caption{Plot of $-{\cal B}_{\eta}$ for the superposition of two photons as
a function of ${\cal J}$ and $\eta$. The maximum violation is $-{\cal
B}_{\eta}=2.07$, which is obtained when ${\cal J}=0.45$ and $\eta = 1$.}
\label{f:TwoPh3D}
\end{figure}

A two-photon state which can be easily produced is the one obtained when
two single-photon states impinge simultaneously in a balanced beam splitter:
in this case the output state is given by
\begin{equation}
\ket{\Omega} = \frac{1}{\sqrt{2}} \big(\ket{2}\ket{0} +
\ket{0}\ket{2}\big)
\end{equation}
and the corresponding Wigner function reads as follows:
\begin{align}
W[\Omega](z,w) =& \:\:
\frac{4}{\pi^2}\,\exp\left\{-2|z|^2-2|w|^2\right\} \nonumber \\
&\times \Big[ 1 - 4\Big(|z|^2 + |w|^2-\big|z^2 - {w^*}^2\big|^2 \Big) \Big]\,,
\end{align}
Now, the  functions defined in Eqs.~(\ref{DefFuns}) are given by 
\begin{align}
{\cal I}_{\eta}(\alpha,\beta) =& 
e^{-\eta(|\alpha|^2-|\beta|^2)}\nonumber\\
&\times\Bigg\{ (1 - \eta)\big[ 1 - \eta 
+ \eta^2 \big( |\alpha|^2 + |\beta|^2 \big)\big]\nonumber\\
&\hspace{1.5cm}- \frac{\eta^4}{4} \big|\alpha^2 - {\cbeta}^2\big|^2
\Bigg\}\,,\\
{\cal G}_{\eta}(\alpha) =& {\cal Y}_{\eta}(\alpha) =
e^{-\eta|\alpha|^2}\,\Bigg[
1 - \eta +\frac{\eta^2}{2} \nonumber\\
&\hspace{1.5cm}+\eta^2 (1 - \eta) |\alpha|^2
+ \frac{\eta^4}{4}|\alpha|^4 \Bigg] \,.\label{G:Y:Omega}
\end{align}
The parametrization which maximizes the violation of the inequality $|{\cal
B}_{\eta}| \le 2$ is $\alpha = \beta = 0$ and $\alpha' = {\beta'}^* =
\sqrt{2}\, e^{i\pi/4} {\cal J}$ and, as for the Bell states, has been
obtained by means of a semi-analytical analysis.  As it is apparent from
Fig.~\ref{f:TwoPh3D}, in this case the violation is quite smaller than the
previous ones: the maximum violation ${\cal B}_{\eta} = -2.07$ is achieved
when ${\cal J} = 0.45$ and $\eta = 1$. Also the threshold for violation
on the quantum efficiency is higher: for $\eta < 92 \%$ we have $|{\cal
B}_\eta| < 2$.
\section{Nonlocality of the twin beam}\label{s:TWB}
The twin-beam state (TWB) of radiation 
$$|r\rangle= \frac{1}{\sqrt{\cosh r}} \sum_{n=0}^\infty \tanh^n r
\: |n\rangle \otimes |n\rangle$$ 
may be produced by spontaneous downconversion in a nonlinear crystal.
TWB is described by the Wigner function
\begin{align}
W_{r}(z,w) =
\frac{4}{\pi^2}\exp\{
-& 2 A (|z|^2+|w|^2) \nonumber\\
&+  2 B (z w + \cz \cw)\}\,,
\label{twb:wig}
\end{align}
with $A \equiv A(r) = \cosh(2 r)$ and $B \equiv B(r) = \sinh (2 r)$, $r$
being the so-called squeezing parameter of the TWB. Since $W_{r}$ and the Wigner
functions of the POVM (\ref{povmhk}) are Gaussian, it is quite simple
to evaluate ${\cal I}_{\eta}(\alpha, \beta)$, ${\cal G}_{\eta}(\alpha)$, and
${\cal Y}_{\eta}(\beta)$ of the correlation function (\ref{E:eta}) and, then,
${\cal B}_{\eta}$; we have
\begin{align}
{\cal I}_{\eta} (\alpha,\beta) = 
\frac{4 {\cal M}_{\eta}(r)}{\eta^2}\,\exp\big\{
-& \widetilde{F}_{\eta} \, (|\alpha|^2 + |\beta|^2)\nonumber\\
&+\widetilde{H}_{\eta} \, (\alpha\beta + \calpha\cbeta)
\big\}
\end{align}
with
\begin{align}
&\widetilde{F}_{\eta}
\equiv\widetilde{F}_{\eta}(r)=
\Delta_{\eta}-(2 A + \Delta_\eta)\,{\cal M}_{\eta}(r)\\
&\widetilde{H}_{\eta} \equiv
\widetilde{H}_{\eta}(r)=  2 B {\cal M}_{\eta}(r)\\
&{\cal M}_{\eta}(r)
= \frac{\Delta_{\eta}^2}{4(A^2 - B^2) + 4 A \Delta_{\eta} +
\Delta_{\eta}^2 }\,,
\end{align}
and
\begin{align}
{\cal G}_{\eta}(\alpha) =& {\cal Y}_{\eta}(\alpha) =
\frac{2 \Delta_{\eta}}{2(A^2 - B^2) + A \Delta_{\eta}}\nonumber\\
& \times
\exp\left\{ -\frac{2 \Delta_{\eta} }
{ 2(A^2 - B^2) + A \Delta_{\eta} }\,|\alpha|^{2} \right\} 
\end{align}
\par
\begin{figure}[h]
\vspace{2.5cm}
\setlength{\unitlength}{.5cm} 
\centerline{%
\begin{picture}(0,0) 
\put(0,0){\makebox(0,0)[c]{\epsfxsize=12\unitlength\epsffile{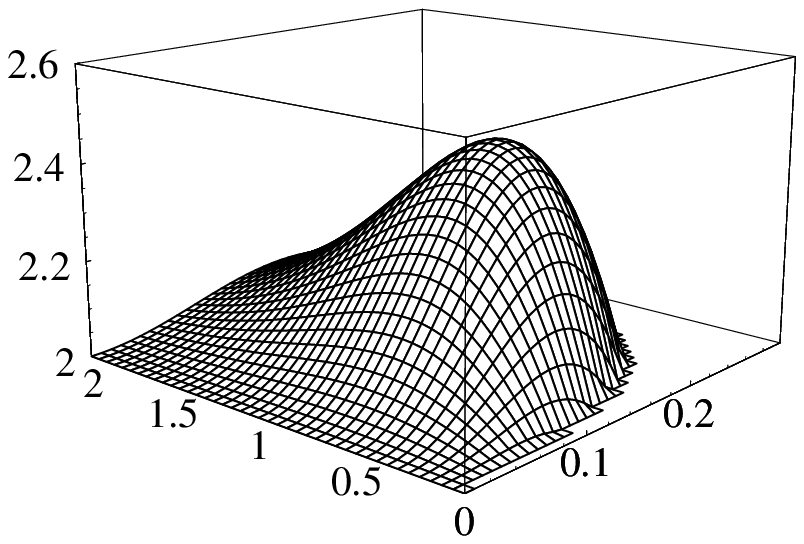}}}
\put(-7.3,1){${\cal B}_{\eta}$}
\put(4,-3.5){${\cal J}$}
\put(-3.5,-3.5){$r$}
\end{picture}
}
\vspace{2cm}
\caption{Plot of ${\cal B}_{\eta}$ for a TWB as a function of ${\cal J}$
and the TWB squeezing parameter $r$ in the case of ideal (i.e., $\eta
= 1$) on/off photodetection. The maximum violation is ${\cal
B}_{\eta}=2.45$, which is obtained when ${\cal J}=0.16$ and $r=0.74$.}
\label{f:3D}
\end{figure}

In order to study Eq.~(\ref{B:param}), we consider the parametrization
$\alpha = -\beta = {\cal J}$ and $\alpha' = -\beta' = -\sqrt{11}{\cal J}$
(as in the case of the Bell states, more details are given in
Sec.~\ref{s:choice}). The parametrization was chosen after a
semi-analytical
analysis and maximizes the violation of the Bell's inequality (for
$\eta=1$). In Fig.~\ref{f:3D} we plot ${\cal B}_{\eta}$ for $\eta=1$:
as one can see the inequality $|{\cal B}_{\eta}|\leq 2$ is violated for a
wide range of parameters, and the maximum violation (${\cal
B}_{\eta}=2.45$) is achieved when ${\cal J}=0.16$ and $r=0.74$.
\par
\begin{figure}[h!]
\vspace{2cm}
\setlength{\unitlength}{.5cm} 
\centerline{%
\begin{picture}(0,0) 
\put(0,0){\makebox(0,0)[c]{\epsfxsize=12\unitlength\epsffile{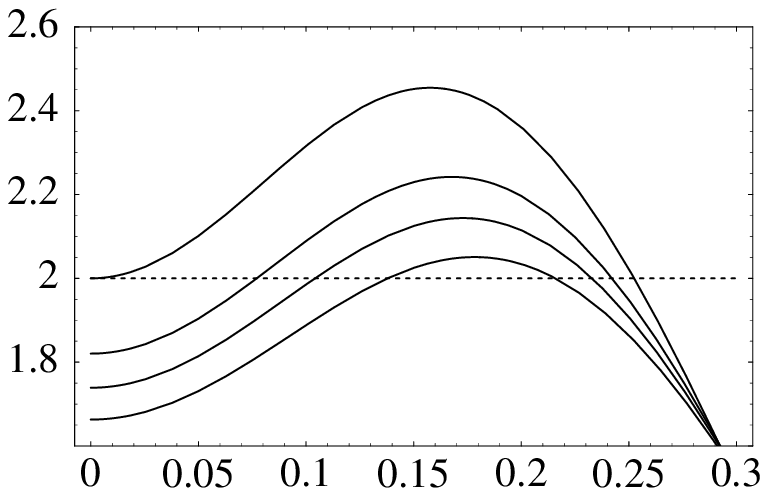}}}
\put(-7,0){${\cal B}_{\eta}$}
\put(0,-4.5){${\cal J}$}
\end{picture}
}
\vspace{2cm}
\caption{Plot of ${\cal B}_{\eta}$ for a TWB as a function of ${\cal J}$
with $r=0.74$ for different values of $\eta$: from top to bottom
$\eta=1.0$, $0.9$, $0.85$, and $0.80$.} \label{f:eta}
\end{figure}

The effect of non-unit efficiency in the detection stage is to reduce the
the violation; this is shown in Fig.~\ref{f:eta}, where we plot ${\cal
B}_{\eta}$ as a function of ${\cal J}$ with $r=0.74$ for different values
of the quantum efficiency. Note that though the violation in the ideal
case, i.e., $\eta=1$, is smaller than for the Bell states, the TWBs
are more robust when one takes into account non-unit quantum
efficiency.  Comparison between Figs.~\ref{f:PsiPhi} and \ref{f:eta} shows
that for $\eta=0.8$ we have a region of ${\cal J}$ values for which ${\cal
B}_{\eta}>2$ in the case of the TWB, whereas there is no violation for
the Bell states. Our parametrization maximize the
violation when $\eta = 1$: in this way $|{\cal B}_{\eta}|\le 2$ when $\eta
< 0.77$ and $r=0.74$. Using different values of $\alpha$, $\beta$,
$\alpha'$, and $\beta'$ (which, now, depend on $\eta$ and the squeezing
parameter $r$), one can extend the violation to lower detection efficiency
\cite{bana:pra:02}. In Sec.~\ref{s:choice} we will draw some remark about
the choice of the parametrization.
\section{Effect of dark counts}\label{s:DarkC}
\begin{figure}[h]
\vspace{4cm}
\setlength{\unitlength}{.35cm} 
\centerline{%
\begin{picture}(0,0) 
\put(0,0){\makebox(0,0)[c]{\epsfxsize=15\unitlength\epsffile{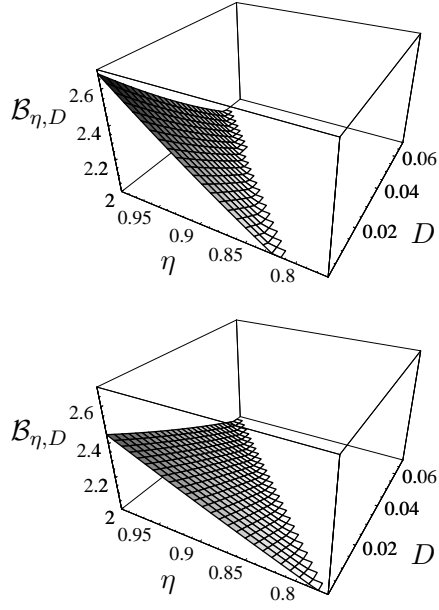}}}
\put(-9.2,7){${\cal B}_{\eta,D}$}
\put(-3.5,1.3){$\eta$}
\put(6,2.3){${D}$}
\put(-9.2,-5.2){${\cal B}_{\eta,D}$}
\put(-3.5,-10.8){${\eta}$}
\put(6,-9.9){$D$}
\end{picture}
}
\vspace{4cm}
\caption{Plot of ${\cal B}_{\eta,D}(\alpha,\beta,\alpha',\beta')$
for the Bell states $\ket{\Phi_\pm}$ (upper plot) and the TWB (lower plot).
We set ${\cal J}=0.17$ for the Bell states, ${\cal J}=0.16$ and $r=0.74$
for the TWB, and used the parametrizations introduced in
Secs.~\ref{s:bbasis} and \ref{s:TWB}, respectively.}\label{f:DarkC}
\end{figure}

In the previous Sections we studied the nonlocality of Bell-like states 
and of the TWB. We took into account the quantum efficiency $\eta$ 
and ignored the effects of dark counts: this is a quite good 
approximation, since, at optical frequencies, dark counts may often 
be neglected. However, there are situations in which the effect
of dark counts cannot be ignored. In this cases, we can add to our 
analysis the effect of the dark counts using Eqs.~(\ref{D2eta}). 
In Fig.~\ref{f:DarkC} we plot ${\cal B}_{\eta,D}$ for the Bell states 
$\ket{\Phi_\pm}$ and the TWB: as on may expect, the violation is reduced.
\par
When the number of dark counts is small we can expand the POVMs 
(\ref{povm1}) and (\ref{povmhk}) up to first order, arriving at 
\begin{align}
&\Pi_0^{(\eta,D)} (\alpha )=
\left(1-D - \eta D \partial_\eta \right)
\Pi_0^{(\eta,0)} (\alpha ) \\ 
&\Pi_{00}^{(\eta,D)} (\alpha,\beta) = 
\left(1- 2 D - \eta D \partial_\eta \right)
\Pi_0^{(\eta,0)} (\alpha,\beta) 
\label{FirstOrder}\;.
\end{align}
Now, using Eq.~(\ref{FirstOrder}) one can express the correlation functions
in two equivalent forms as follows
\begin{align}
E_{\eta,D} (\alpha,\beta) 
&= \left(1- 2 D - \eta D \partial_\eta \right)
E_{\eta,0} (\alpha,\beta) \nonumber\\
&\hspace{5mm}+2 D \left[{\cal G}_{\eta,0} (\alpha) 
+ {\cal Y}_{\eta,0} (\beta)\right] \\
&= \left(1- D - \eta D \partial_\eta \right)
E_{\eta,0} (\alpha,\beta) \nonumber \\ 
&\hspace{5mm}- 4 D {\cal I}_{\eta,0} (\alpha,\beta) 
\label{EsmallD}\;,
\end{align}
which, in turn, can be used to express the Bell parameter, 
as follows
\begin{align}
B_{\eta,D} = & 
\left(1- 2 D - \eta D \partial_\eta \right)
B_{\eta,0} (\alpha,\beta)\nonumber\\
& + \left. 4 D \left[{\cal G}_{\eta,0} (\alpha) \right.
+ {\cal Y}_{\eta,0} (\beta)\right]\,.
\label{BsmallD}\;
\end{align}
\section{Nonlocality of the de-Gaussified twin beam}\label{s:IPS}
The de-Gaussification of a TWB can be achieved by subtracting
photons from both modes \cite{ips:PRA:67,opatr:PRA:61,coch:PRA:65}.
In Ref.~\cite{ips:PRA:67} we referred to this process as to inconclusive
photon subtraction (IPS) and showed that the resulting state, the IPS
state, can be used to enhance the
teleportation fidelity of coherent states for a wide range of the
experimental parameters. Moreover, 
in Ref.~\cite{ips:PRA:70}, we have shown that, in the absence of any noise
during the transmission stage, the IPS state has nonlocal correlations
larger than those of the TWB irrespective of the IPS quantum efficiency
(see also Refs.~\cite{nha:PRL:93,garcia:PRL:93}).
\par
\begin{figure}
\includegraphics[scale=.8]{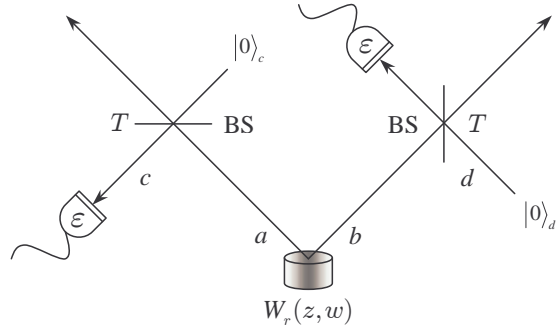}
\vspace{-.3cm}
\caption{\label{f:IPS:scheme} Scheme of the IPS process.}
\end{figure}

First of all we briefly recall the IPS process, whose 
scheme is sketched in Fig.~\ref{f:IPS:scheme}. The two
modes, $a$ and $b$, of the TWB are mixed with the vacuum (modes
$c$ and $d$, respectively) at two unbalanced beam splitters (BS)
with equal transmissivity; the  modes $c$ and
$d$ are then detected by avalanche photodetectors (APDs) with equal
efficiency, which can only discriminate the presence of
radiation from the vacuum: the IPS state is obtained when the two
detectors jointly click. When the
input state, namely the state arriving at the two beam splitters, is
the TWB of Eq.~(\ref{twb:wig}), the state produced
by the IPS process reads as follows (see Ref.~\cite{ips:PRA:70} for
details)
\begin{equation}\label{ips:wigner}
W_{r,T,\varepsilon}^{\rm (IPS)}(z,w) =
\frac{4}{\pi^2\,p_{11}(r,T,\varepsilon)} 
\sum_{k=1}^4 {\cal C}_k\,
W_{r,T,\varepsilon}^{(k)}(z,w)\,,
\end{equation}
where
\begin{equation}\label{ips:probability}
p_{11}(r,T,\varepsilon) =
\sum_{k=1}^4 \frac{{\cal C}_k}{F_k G_k - H_k^2}\,
\end{equation}
is the probability of a click in both the APDs. In Eqs.~(\ref{ips:wigner})
and (\ref{ips:probability}) we introduced
\begin{equation}
{\cal C}_k \equiv {\cal C}_k(r,T,\varepsilon)=
\frac{4\,C_k}
{x_k y_k - 4 \wtB^2 (1-T)^2}\,,
\end{equation}
and defined
\begin{align}
W_{r,T,\varepsilon}^{(k)}(z,w) =
\exp\{ -&F_k |z|^2 - G_k |w|^2\nonumber\\ 
&+ H_k (z w + \cz \cw)\}\,,
\end{align}
where $F_k = (b - f_k)$, $G_k = (b - g_k)$, $H_k = (2 B T + h_k)$,
$C_k\equiv C_k(\varepsilon)$ with
$C_1 = 1$, $C_2 = C_3 = -2(2-\varepsilon)^{-1}$,
$C_4 = 4(2-\varepsilon)^{-2}$; $x_k \equiv x_k(r,T,\varepsilon)$ and
$y_k \equiv y_k(r,T,\varepsilon)$ are
\begin{align}
&x_1 = x_3 = y_1 = y_2 = a \nonumber\\
&x_2 = x_4 = y_3 = y_4 = a +2\varepsilon (2-\varepsilon)^{-1} \nonumber
\end{align}
with $a \equiv a(r,T) = 2 [\wtA (1-T) + T]$,
$b \equiv b(r,T) = 2 [\wtA T + (1-T)]$; finally, $f_k$, $g_k$,
and $h_k$ depend on $r$, $T$ and $\varepsilon$ and are given by
\begin{align}
f_k &= \calN_k
\, [x_k \wtB^2\nonumber\\
&\hspace{1.5cm} + 4 \wtB^2 (1-\wtA) (1-T) + y_k (1-\wtA)^2]\,,\\
g_k &= \calN_k\, [x_k (1-\wtA)^2\nonumber\\
&\hspace{1.5cm} + 4 \wtB^2 (1-\wtA) (1-T) + y_k \wtB^2]\,,\\
h_k &= \calN_k
\, \{(x_k + y_k) \wtB (1-\wtA)\nonumber\\
&\hspace{1.5cm}+ 2 \wtB [\wtB^2 + (1-\wtA)^2] (1-T)\}\,,\\
\calN_k &\equiv {\calN_k}(r,T,\varepsilon) = {\displaystyle
\frac{4 T\, (1-T)}{x_k y_k - 4 \wtB^2 (1-T)^2}\,.
}
\end{align}
The state given in Eq.~(\ref{ips:wigner}) is no longer a Gaussian state
and, in the following, we will use the measurement described above in order
to test its nonlocality.
Nonlocal properties of the IPS state (\ref{ips:wigner}) have been
investigated in Refs.~\cite{ips:PRA:70,ips:noise} by means of other kinds
of nonlocality tests. In particular, Ref.~\cite{ips:noise} addressed the
presence of noise during the propagation and detection stages, showing that
the IPS process onto TWBs is a quite robust method to enhance their
nonlocal correlations especially in the low energy (i.e., small $r$)
regime.
\par
In the case of the state (\ref{ips:wigner}), the correlation function
(\ref{E:eta}) reads (for the sake of simplicity we do not write explicitly
the dependence on $r$, $T$ and $\varepsilon$)
\begin{align}
E_{\eta}(\alpha,\beta)=&
\frac{1}{p_{11}(r,T,\varepsilon)}
\sum_{k=1}^4 {\cal C}_k\,
\big\{ 1 + 4\, {\cal I}^{(k)}_{\eta}(\alpha, \beta)
\nonumber\\
&- 2 \big[{\cal G}^{(k)}_{\eta}(\alpha) +
 {\cal Y}^{(k)}_{\eta}(\beta)\big]\big\}\,,
\end{align}
where
\begin{align}
{\cal I}^{(k)}_{\eta} (\alpha,\beta) =&
\frac{4 {\cal M}_{\eta}^{(k)}(r,T,\varepsilon)}{\eta^2}\,
\exp\big\{
-\widetilde{G}_{\eta}^{(k)} \, |\alpha|^2\nonumber\\
&-\widetilde{F}_{\eta}^{(k)} \, |\beta|^2
+\widetilde{H}_{\eta}^{(k)} \, (\alpha\beta + \calpha\cbeta)
\big\}\,,
\end{align}
with $\widetilde{F}_{\eta}^{(k)}
\equiv\widetilde{F}_{\eta}^{(k)}(r,T,\varepsilon)$,
$\widetilde{G}_{\eta}^{(k)}
\equiv\widetilde{G}_{\eta}^{(k)}(r,T,\varepsilon)$, and
$\widetilde{H}_{\eta}^{(k)}
\equiv\widetilde{H}_{\eta}^{(k)}(r,T,\varepsilon)$ given by
\begin{align}
&\widetilde{F}_{\eta}^{(k)} =
\Delta_{\eta}-(F_k+\Delta_\eta)\,{\cal M}_{\eta}^{(k)}(r,T,\varepsilon)\,,\\
&\widetilde{G}_{\eta}^{(k)} = 
\Delta_{\eta}-(G_k+\Delta_\eta)\,{\cal M}_{\eta}^{(k)}(r,T,\varepsilon)\,,\\
&\widetilde{H}_{\eta}^{(k)} =
H_k\,{\cal M}_{\eta}^{(k)}(r,T,\varepsilon)\,,\\
&{\cal M}_{\eta}^{(k)}(r,T,\varepsilon)
= \frac{\Delta_{\eta}^2}{(F_k+\Delta_{\eta})(G_k+\Delta_{\eta}) - H_k^2}\,,
\end{align}
respectively, and
\begin{align}
{\cal G}^{(k)}_{\eta}(\alpha) =&
\frac{4 \Delta_{\eta}}{[G_k\,(F_k + \Delta_{\eta}) -
H_k^2]\,\eta}\nonumber\\
&\times \exp\left\{
-\frac{(F_k G_k - H_k^2)\,\Delta_{\eta}}{G_k\,(F_k +
\Delta_{\eta})-H_k^2}\,|\alpha|^{2} \right\}\,, \\
%
{\cal Y}^{(k)}_{\eta}(\beta) =&
\frac{4 \Delta_{\eta}}{[F_k\,(G_k + \Delta_{\eta}) -
H_k^2]\,\eta}\nonumber\\
&\times \exp\left\{
-\frac{(F_k G_k - H_k^2)\,\Delta_{\eta}}{F_k\,(G_k +
\Delta_{\eta})-H_k^2}\,|\beta|^{2} \right\}\,.
\end{align}
\par
\begin{figure}[h]
\vspace{2.5cm}
\setlength{\unitlength}{.5cm} 
\centerline{%
\begin{picture}(0,0) 
\put(0,0){\makebox(0,0)[c]{\epsfxsize=12\unitlength\epsffile{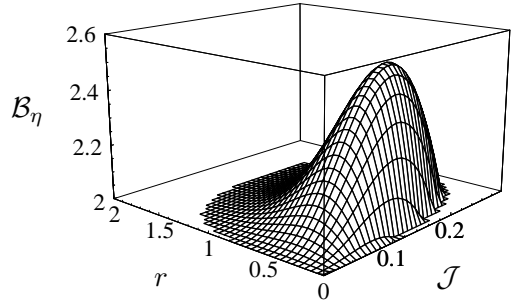}}}
\put(-7.3,1){${\cal B}_{\eta}$}
\put(4,-3.5){${\cal J}$}
\put(-3.5,-3.5){$r$}
\end{picture}
}
\vspace{2cm}
\caption{Plot of ${\cal B}_{\eta}$ for the IPS state with $T = 0.9999$ and
$\varepsilon = 1$
as a function of ${\cal J}$ and the TWB squeezing parameter $r$ in the case
of ideal (i.e., $\eta = 1$) on/off photodetection. The maximum
violation is ${\cal B}_{\eta}=2.53$, which is obtained when ${\cal J}=0.16$
and $r=0.39$.} \label{f:IPS3D}
\end{figure}
\begin{figure}[h]
\vspace{2cm}
\setlength{\unitlength}{.5cm} 
\centerline{%
\begin{picture}(0,0) 
\put(0,0){\makebox(0,0)[c]{\epsfxsize=12\unitlength\epsffile{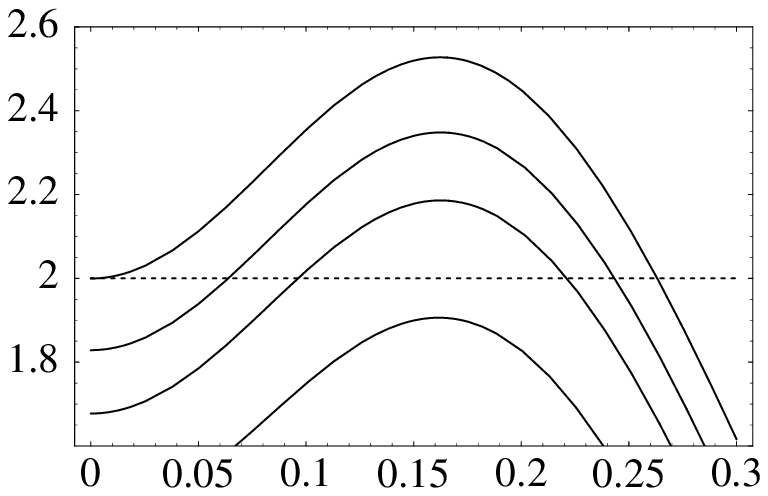}}}
\put(-7,0){${\cal B}_{\eta}$}
\put(0,-4.5){${\cal J}$}
\end{picture}
}
\vspace{2cm}
\caption{Plot of ${\cal B}_{\eta}$ for the IPS state as a function of
${\cal J}$ with $r=0.39$ for different values of $T$ and $\varepsilon = 1$
in the ideal case (i.e., $\eta = 1$): from top to bottom
$T=0.9999$, $0.95$, $0.90$, and $0.80$.} \label{f:IPStau}
\end{figure}

In order to investigate the nonlocality of the IPS by means of
Eq.~(\ref{B:param}), we choose the same parametrization as in
Sec.~\ref{s:TWB}.
The results are showed in Figs.~\ref{f:IPS3D} and \ref{f:IPStau} for $\eta
= 1$ and $\varepsilon = 1$: we can see that the IPS enhances the violation of the inequality
$|{\cal B}_{\eta}|\le 2$ for small values of $r$ (see also
Refs.~\cite{ips:PRA:67, ips:PRA:70, ips:noise}). Moreover, as one may expect,
the maximum of violation is achieved as $T \to 1$, whereas decreasing the
effective transmission of the IPS process, one has that the inequality becomes
satisfied for all the values of $r$, as we can see in Fig.~\ref{f:IPStau}
for $\tau = 0.8$.
\par
\begin{figure}[h]
\vspace{2cm}
\setlength{\unitlength}{.5cm} 
\centerline{%
\begin{picture}(0,0) 
\put(0,0){\makebox(0,0)[c]{\epsfxsize=12\unitlength\epsffile{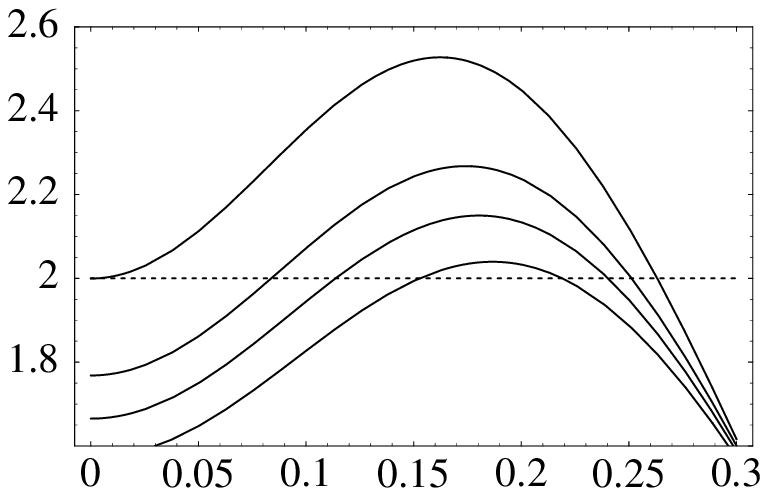}}}
\put(-7,0){${\cal B}_{\eta}$}
\put(0,-4.5){${\cal J}$}
\end{picture}
}
\vspace{2cm}
\caption{Plot of ${\cal B}_{\eta}$ for the IPS state as a function of
${\cal J}$ with $r=0.39$, $T = 0.9999$, $\varepsilon = 1$, 
and for different values of
$\eta$: from top to bottom $\eta=1.0$, $0.9$, $0.85$, and $0.8$.}
\label{f:IPSeta}
\end{figure}
\begin{figure}[h]
\vspace{2.5cm}
\setlength{\unitlength}{.5cm} 
\centerline{%
\begin{picture}(0,0) 
\put(0,0){\makebox(0,0)[c]{\epsfxsize=12\unitlength\epsffile{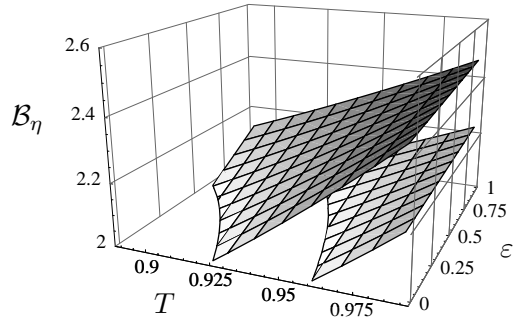}}}
\put(-7.3,1){${\cal B}_{\eta}$}
\put(5.7,-2.5){$\varepsilon$}
\put(-3.5,-4){$T$}
\end{picture}
}
\vspace{2.2cm}
\caption{Plot of ${\cal B}_{\eta}$ for the IPS state as a function $T$ and
$\varepsilon$ with ${\cal J} = 0.16$, $r=0.39$, and, from top to bottom,
$\eta = 0.99$, and $0.90$. The main effect on ${\cal B}_{\eta}$ is due to
the transmissivity $T$.} \label{f:IPSTauEta}
\end{figure}

In Fig.~\ref{f:IPSeta} we plot ${\cal B}_{\eta}$ for the IPS
with $T = 0.9999$, $\varepsilon = 1$ and different $\eta$.
As for the TWB, we can have
violation of the Bell's inequality also for detection efficiencies near to
$80 \%$. As for the Bell states and the TWB, a $\eta$- and $r$-dependent
choice of the parameters in Eq.~(\ref{B:param}) can improve this result.
The effect on a non-unit $\varepsilon$ is studied in
Fig.~\ref{f:IPSTauEta}, where we plot ${\cal B}_\eta$ as a function of $T$
and $\varepsilon$ and fixed values of the other involved parameters.
We can see that the main effect on the Bell parameter is due to the
transmissivity $T$.
\par
The presence of dark counts at the detection stage can be taken into
account using Eqs.~(\ref{D2eta}): since the results are similar to those
of the Bells states and the TWB presented in Sec.~\ref{s:DarkC}, we do
not report them explicitly.
\section{Choice of the parametrization}\label{s:choice}
\begin{figure}[h]
\vspace{5cm}
\setlength{\unitlength}{.4cm} 
\centerline{%
\begin{picture}(0,0) 
\put(0,0){\makebox(0,0)[c]{\epsfxsize=12\unitlength\epsffile{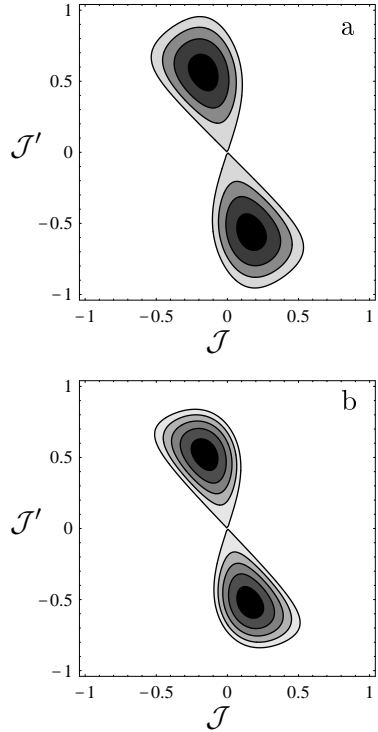}}}
\put(-6.5,6.5){${\cal J}'$}
\put(0,0){${\cal J}$}
\put(-6.5,-6){${\cal J}'$}
\put(0,-12.5){${\cal J}$}
\put(4.5,10.5){a}
\put(4.5,-2){b}
\end{picture}
}
\vspace{5cm}
\caption{Plots of $|{\cal B}_{\eta}|$ for (a) the states $\ket{\Psi_{\pm}}$
and (b) the TWB with $r=0.74$ and $\eta=1$.  The darker is the region, the
bigger is the violation of the Bell's inequality. In the white region
$|{\cal B}_{\eta}|\le 2$. ${\cal J}$ and ${\cal J}'$ refer to the
particular parametrization of ${\cal B}_{\eta}$, see the text for details.}
\label{f:PhiTwb}
\end{figure}

In this Section we draw some remark about the choice of the parametrization
used in the investigation of the Bell parameter ${\cal B}_{\eta}$. Numerical
analysis has shown that, in the case of state $\ket{\Psi_{+}}$ and in the
presence of non-unit quantum efficiency, the
maximal violation of the Bell's inequality for the displaced on/off test
is achieved by choosing $\alpha$, $\alpha'$, $\beta$, and $\beta'$ as
complex parameters \cite{bana:pra:02}. On the other hand,
here we addressed only {\em real} parametrization for the
Bell's parameter ${\cal B}$ given in Eq.~(\ref{B:param}), and,
in particular, we take $\alpha = -\beta = {\cal
J}\in\mathbb{R}$, and $\alpha' = -\beta' = {\cal J}'\in\mathbb{R}$ for
the states $\ket{\Psi_{+}}$, $\ket{\Phi_{+}}$, the TWB and the IPS state,
while we put $\alpha = \beta = {\cal J}\in\mathbb{R}$ and
$\alpha' = \beta' = {\cal J}'\in\mathbb{R}$ for the states
$\ket{\Psi_{-}}$ and $\ket{\Phi_{-}}$.
In Fig.~\ref{f:PhiTwb} we plot $|{\cal B}_{\eta}|$ as a 
function of ${\cal J}$ and ${\cal J}'$ in the ideal case (i.e.,
$\eta=1$) for (a) the states $\ket{\Psi_{\pm}}$ and (b) the TWB with
$r=0.74$, which maximizes the violation. The results for the other states
are similar. In both the plots, the darker is the region, the bigger is the
violation (the white region refers to $|{\cal B}_{\eta}|\le 2$). As one can see,
there is a symmetry with respect to the origin, which implies that the best
parametrization has the form ${\cal J}' = -\kappa {\cal J}$, with
$\kappa\in\mathbb{R}$, $\kappa>0$. Furthermore, for all the considered
states, the numerical analysis shows that a good choice for $\kappa$ is
$\kappa=\sqrt{11}$, which is an approximation of the actual value.
\par
\begin{figure}[h]
\vspace{4cm}
\setlength{\unitlength}{.6cm} 
\centerline{%
\begin{picture}(0,0) 
\put(0,0){\makebox(0,0)[c]{\epsfxsize=13\unitlength\epsffile{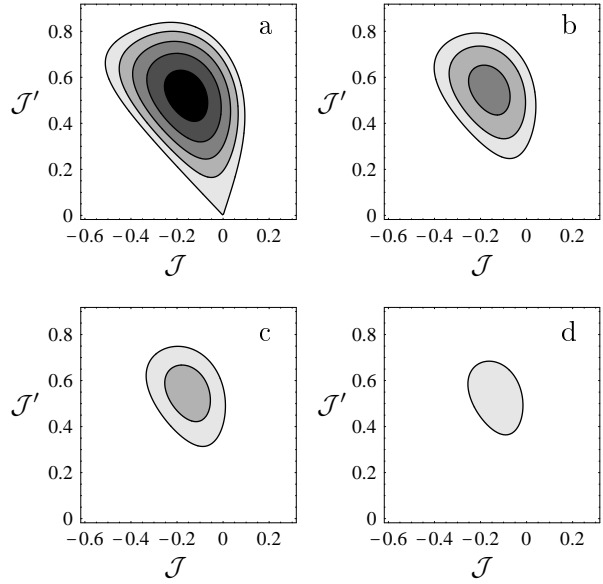}}}
\put(-6.9,3.7){\footnotesize ${\cal J}'$} 
\put(-3.5,0.1){\footnotesize ${\cal J}$}  
\put(-0.1,3.7){\footnotesize ${\cal J}'$} 
\put(3.3,0.1){\footnotesize ${\cal J}$}   
\put(-6.9,-3){\footnotesize ${\cal J}'$}  
\put(-3.5,-6.6){\footnotesize ${\cal J}$} 
\put(-0.1,-3){\footnotesize ${\cal J}'$}  
\put(3.3,-6.6){\footnotesize ${\cal J}$}  
\put(-1.4,5.4){a}
\put(5.3,5.4){b}
\put(-1.4,-1.4){c}
\put(5.3,-1.4){d}
\end{picture}
}
\vspace{4cm}
\caption{Plots of ${\cal B}_{\eta}$ in the case of TWB for different $\eta$.
The TWB parameter $r$ is chosen in order to maximize the violation of the Bell's
inequality as $\eta$ varies. We put
(a) $\eta = 1$,    $r=0.74$ (${\cal B}_{\eta, {\rm max}}=2.45$),
(b) $\eta = 0.9$,  $r=0.67$ (${\cal B}_{\eta, {\rm max}}=2.24$),
(c) $\eta = 0.85$, $r=0.60$ (${\cal B}_{\eta, {\rm max}}=2.15$), and
(d) $\eta = 0.8$,  $r=0.49$ (${\cal B}_{\eta, {\rm max}}=2.07$).
${\cal J}$ and ${\cal J}'$ refer to the
particular parametrization of ${\cal B}_{\eta}$, see the text for details.
Since there is symmetry with respect to the origin, we show only the
region ${\cal J}'\ge 0$.}
\label{f:TwbParamEta}
\end{figure}

In Fig.~\ref{f:TwbParamEta} the effect of $\eta$ is taken into account:
Since the results for the Bell's states, the TWB and the IPS state are
similar, we only address the TWB case: there is still the symmetry with
respect to the origin, but a thorough numerical investigation shows that
the maximum of ${\cal B}_{\eta}$, and, then, $\kappa$ depend on both $\eta$ and
$r$.
\par
Notice that we have considered real values for the parameters. 
It can be shown numerically \cite{bana:pra:02} that for decreasing 
$\eta$ a complex parametrization leads to a slight
improvement.
\section{Bell's inequality, POVMs and maximum violation}
\label{s:cirel}
In this Section we address the maximal violation of the Bell inequality
that is achievable by using non-projective measurements.
\par
Let us consider two systems, ${\sf A}$ and ${\sf B}$, and the generic POVM
$\{\Pi_0(\zeta), \Pi_1(\zeta) \}$, depending on the complex parameter
$\zeta$, such that $\Pi_1(\zeta) = \iid - \Pi_0(\zeta)$. We define the
observables
\begin{equation}
\Oop_{\sf k}(\zeta) = \Pi_1(\zeta) - \Pi_0(\zeta) = \iid - 2\,\Pi_0(\zeta)
\end{equation}
acting on system ${\sf k}={\sf A}, {\sf B}$, respectively (we are using the
same POVM for both the systems). Furthermore, we assume that $\Oop_{\sf
A}(\zeta)$ and $\Oop_{\sf B}(\zeta)$ have spectra included in the interval
$[-1,+1]$ \cite{cirelsonB}. Now we introduce the Bell operator \cite{CHSH}
\begin{align}
\Bop =& \Oop_{\sf A}(\alpha)\otimes \Oop_{\sf B}(\beta) +
 \Oop_{\sf A}(\alpha')\otimes \Oop_{\sf B}(\beta)\nonumber\\
 &+
 \Oop_{\sf A}(\alpha)\otimes \Oop_{\sf B}(\beta') -
 \Oop_{\sf A}(\alpha')\otimes \Oop_{\sf B}(\beta')\,,
\end{align}
which has the property \cite{cirelsonB}
\begin{align}
\Bop \le \frac{1}{\sqrt{2}} \big[&  \Oop_{\sf A}^2(\alpha)\otimes \iid +
\iid \otimes \Oop_{\sf B}^2(\beta) \nonumber\\
& + \Oop_{\sf A}^2(\alpha')\otimes \iid +
\iid \otimes \Oop_{\sf B}^2(\beta') \big]\,.\label{Bmaj}
\end{align}
If $\Pi_1(\zeta)$ and $\Pi_0(\zeta)$ are projectors on orthogonal subspaces,
namely $\Pi_1^2(\zeta) = \Pi_1(\zeta)$, $\Pi_0^2(\zeta) = \Pi_0(\zeta)$ and
$\Pi_0(\zeta)\,\Pi_1(\zeta) = 0$, then
\begin{equation}
 \Oop_{\sf A}^2(\zeta) = \Oop_{\sf B}^2(\xi) = \iid\,,
\end{equation}
and Eq.~(\ref{Bmaj}) leads to
\begin{equation}\label{Cbound}
{\cal B} \le 2\sqrt{2}\,,
\end{equation}
where ${\cal B} = {\rm Tr}[\varrho\, \Bop]$, $\varrho$ being the state of
the system, is the Bell parameter. The bound $2\sqrt{2}$ is usually
known as {\em Cirel'son bound} and is the maximum violation achievable in
the case of a bipartite quantum system \cite{cirelsonB}.
Eq.~(\ref{Cbound}) may be also derived
in a different \cite{landau,peres}:
since the squared Bell operator reads
\begin{equation}
\Bop^2 = 4 \iid +
\left[\Oop_{\sf A}(\alpha),\Oop_{\sf A}(\alpha')\right]\otimes
\left[\Oop_{\sf B}(\beta),\Oop_{\sf B}(\beta')\right]\,,
\end{equation} 
then  using the relation $|| [A,B] || \le 2 || A ||\, || B ||$, 
where $|| A || = {\rm Sup}_{|\psi\rangle} ||A|\psi\rangle||$,
we have ${\cal B}^2 \le 8$,
from which  Eq.~(\ref{Cbound}) follows.
\par
On the other hand, when $\{\Pi_0(\zeta), \Pi_1(\zeta) \}$ is not a
projective measurement a different inequality should be derived. First of
all we note that the observables $\Oop_{\sf k}(\zeta)$ corresponding to the
POVM given in Eq.~(\ref{povm1}) satisfy the hypothesis of the Cirel'son
theorem. In fact, in this case
\begin{equation}
\Oop_{\sf k}(\zeta) = \sum_{n=0}^{\infty}
\left[1-2(1-\eta)^n\right] D(\zeta) \ket{n}\bra{n} D^{\dag}(\zeta)
\end{equation}
and its spectrum $\{ \lambda_n \}$, $\lambda_n = 1- 2 (1-\eta)^n$,
lies in the interval $[-1,+1]$ for $0\le \eta \le 1$ (when $\eta = 1$ the
spectrum reduces to the two points $\{-1,+1\}$). Now, one has
\begin{align}
 \Oop_{\sf A}^2(\zeta) &= \iid - 4\,\Pi_0(\zeta)\,\Pi_1(\zeta) =
\iid - 4\,[\Pi_0(\zeta) - \Pi_0^2(\zeta)]
\nonumber\\
&= \iid - 4\,\Eop_{\sf A}(\zeta)\,,
\end{align}
and, analogously, $\Oop_{\sf B}^2(\xi) = \iid - 4\, \Eop_{\sf B}(\xi)$,
where we defined the operator
\begin{equation}
\Eop_{\sf k}(\zeta) = \Pi_0(\zeta) - \Pi_0^2(\zeta)\,,
\end{equation}
${\sf k} = {\sf A}, {\sf B}$. In this way, from Eq.~(\ref{Bmaj}) follows
\begin{equation}
\Bop \le 2\sqrt{2} \big\{ \iid - [\Eop_{\sf A}(\alpha) +
\Eop_{\sf B}(\beta) + \Eop_{\sf A}(\alpha') +
\Eop_{\sf B}(\beta')]\big\}\,,
\end{equation}
with $\Eop_{\sf A}(\zeta) \equiv \Eop_{\sf A}(\zeta)\otimes \iid$ and
 $\Eop_{\sf B}(\xi) \equiv \iid\otimes \Eop_{\sf B}(\xi)$. Finally
we get
\begin{equation}\label{BnewBound}
{\cal B} \le {\cal B}^{\rm (max)}(\alpha,\beta,\alpha',\beta')\,,
\end{equation}
where we defined
\begin{align}
{\cal B}^{\rm (max)} &\equiv {\cal B}^{\rm
(max)}(\alpha,\beta,\alpha',\beta')\\
&= 2\sqrt{2} \big\{ 1- [{\cal E}_{\sf A}(\alpha) + 
 {\cal E}_{\sf B}(\beta) \nonumber\\
&\hspace{3cm}  + {\cal E}_{\sf A}(\alpha') +
 {\cal E}_{\sf B}(\beta')] \big\}
\end{align}
and ${\cal E}_{\sf A}(\zeta) = {\rm Tr}[\Eop_{\sf A}(\zeta)\otimes
\iid]$ and ${\cal E}_{\sf B}(\xi) = {\rm Tr}[\iid\otimes \Eop_{\sf
B}(\xi)]$.  Now, since ${\cal E}_{\sf k} \ge 0$, ${\sf k} = {\sf A},
{\sf B}$, one has that the bound of the Bell parameter is smaller than the
limit $2\sqrt{2}$, obtained in the case of projective measurements. Notice
that the new bound depends on the parameters $\alpha$, $\beta$, $\alpha'$,
and $\beta'$ of the measurement and on the state under investigation
itself.
\par
\begin{figure}[h]
\vspace{5cm}
\setlength{\unitlength}{.35cm} 
\centerline{%
\begin{picture}(0,0) 
\put(0,0){\makebox(0,0)[c]{\epsfxsize=12.9\unitlength\epsffile{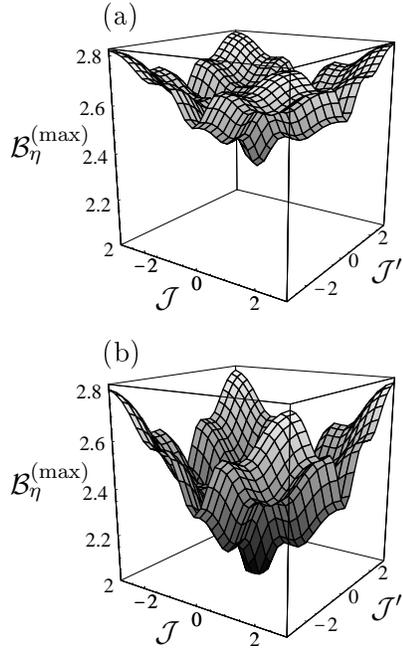}}}
\put(-5,12){(a)}
\put(-8.5,7){${\cal B}^{\rm (max)}_{\eta}$}
\put(-3,1.3){$\cal J$}
\put(5.2,2.3){${\cal J'}$}
\put(-5,-0.8){(b)}
\put(-8.5,-5.8){${\cal B}^{\rm (max)}_{\eta}$}
\put(-3,-11.4){${\cal J}$}
\put(5.2,-10.5){$\cal J'$}
\end{picture}
}
\vspace{4.5cm}
\caption{Plot of ${\cal B}^{\rm (max)}_{\eta}(\alpha,\beta,\alpha',\beta')$
for the Bell states. We set $\alpha=\beta={\cal J}$ and
$\alpha'=\beta'={\cal J'}$ and: (a) $\eta = 0.9$, (b) $\eta = 0.8$.}
\label{f:B3D}
\end{figure}
\begin{figure}[h]
\vspace{2cm}
\setlength{\unitlength}{.45cm} 
\centerline{%
\begin{picture}(0,0) 
\put(0,0){\makebox(0,0)[c]{\epsfxsize=12\unitlength\epsffile{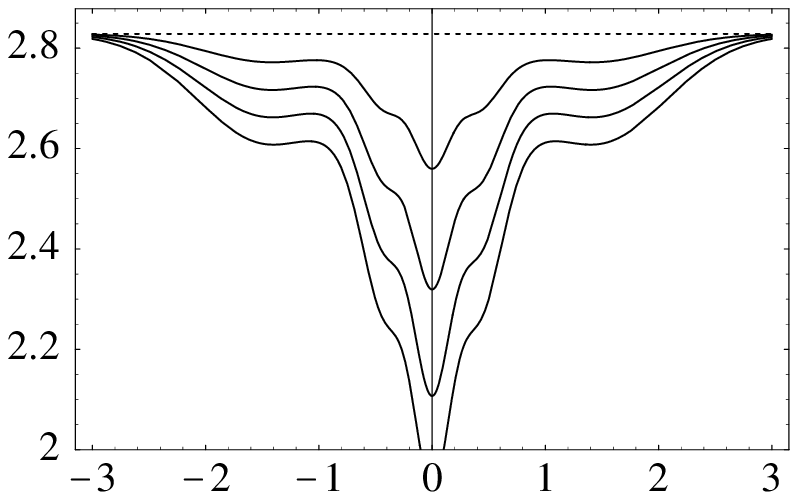}}}
\put(-8,0){${\cal B}^{\rm (max)}_{\eta}$}
\put(0.2,-4.5){$\cal J$}
\end{picture}
}
\vspace{2cm}
\caption{Plots of ${\cal B}^{\rm (max)}_{\eta}(\alpha,\beta,\alpha',\beta')$
in the case of the Bell states
for different values of $\eta$; from top to bottom (solid lines): $\eta =
0.95, 0.9, 0.85$, and $0.8$. The dashed line corresponds to the value
$2\sqrt{2}$ obtained when $\eta = 1$. We set $\alpha = \beta = {\cal J}$ and
$\alpha' = \beta' = \sqrt{11}\,{\cal J}$, which maximize ${\cal B}$.}
\label{f:maxBpar}
\end{figure}
\begin{figure}[h]
\vspace{5cm}
\setlength{\unitlength}{.35cm} 
\centerline{%
\begin{picture}(0,0) 
\put(0,0){\makebox(0,0)[c]{\epsfxsize=12.9\unitlength\epsffile{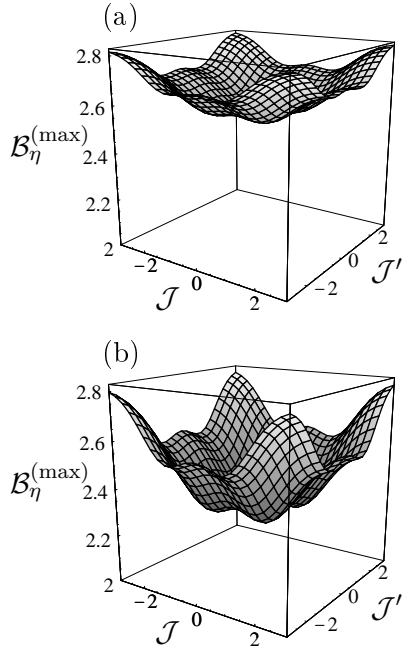}}}
\put(-5,12){(a)}
\put(-8.5,7){${\cal B}^{\rm (max)}_{\eta}$}
\put(-3,1.3){$\cal J$}
\put(5.2,2.3){${\cal J'}$}
\put(-5,-0.8){(b)}
\put(-8.5,-5.8){${\cal B}^{\rm (max)}_{\eta}$}
\put(-3,-11.4){${\cal J}$}
\put(5.2,-10.5){$\cal J'$}
\end{picture}
}
\vspace{4.5cm}
\caption{Plot of ${\cal B}^{\rm (max)}_{\eta}(\alpha,\beta,\alpha',\beta')$
for the TWB with $r=0.74$. We set $\alpha=\beta={\cal J}$ and
$\alpha'=\beta'={\cal J'}$ and: (a) $\eta = 0.9$, (b) $\eta = 0.8$.}
\label{f:BTWB3D}
\end{figure}
\begin{figure}[h]
\vspace{2cm}
\setlength{\unitlength}{.45cm} 
\centerline{%
\begin{picture}(0,0) 
\put(0,0){\makebox(0,0)[c]{\epsfxsize=12\unitlength\epsffile{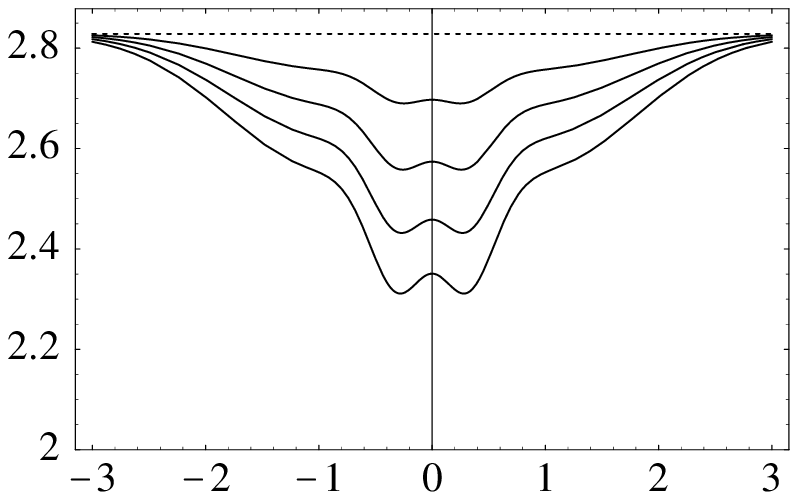}}}
\put(-8,0){${\cal B}^{\rm (max)}_{\eta}$}
\put(0.2,-4.5){$\cal J$}
\end{picture}
}
\vspace{2cm}
\caption{Plots of ${\cal B}^{\rm (max)}_{\eta}(\alpha,\beta,\alpha',\beta')$
in the case of the TWB with $r=0.74$
for different values of $\eta$; from top to bottom (solid lines): $\eta =
0.95, 0.9, 0.85$, and $0.8$. The dashed line corresponds to the value
$2\sqrt{2}$ obtained when $\eta = 1$. We set $\alpha = \beta = {\cal J}$ and
$\alpha' = \beta' = \sqrt{11}\,{\cal J}$, which maximize ${\cal B}$.}
\label{f:maxBparTWB}
\end{figure}
\begin{figure}[h]
\vspace{5cm}
\setlength{\unitlength}{.35cm} 
\centerline{%
\begin{picture}(0,0) 
\put(0,0){\makebox(0,0)[c]{\epsfxsize=12.9\unitlength\epsffile{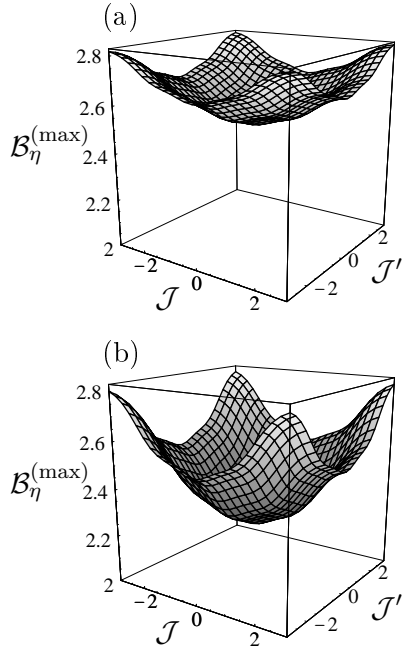}}}
\put(-5,12){(a)}
\put(-8.5,7){${\cal B}^{\rm (max)}_{\eta}$}
\put(-3,1.3){$\cal J$}
\put(5.2,2.3){${\cal J'}$}
\put(-5,-0.8){(b)}
\put(-8.5,-5.8){${\cal B}^{\rm (max)}_{\eta}$}
\put(-3,-11.4){${\cal J}$}
\put(5.2,-10.5){$\cal J'$}
\end{picture}
}
\vspace{4.5cm}
\caption{Plot of ${\cal B}^{\rm (max)}_{\eta}(\alpha,\beta,\alpha',\beta')$
for the IPS state with $r=0.39$, $T=0.9999$, and $\varepsilon = 1$.
We set $\alpha=\beta={\cal J}$ and
$\alpha'=\beta'={\cal J'}$ and: (a) $\eta = 0.9$, (b) $\eta = 0.8$.}
\label{f:BIPS3D}
\end{figure}
\begin{figure}[h]
\vspace{2cm}
\setlength{\unitlength}{.45cm} 
\centerline{%
\begin{picture}(0,0) 
\put(0,0){\makebox(0,0)[c]{\epsfxsize=12\unitlength\epsffile{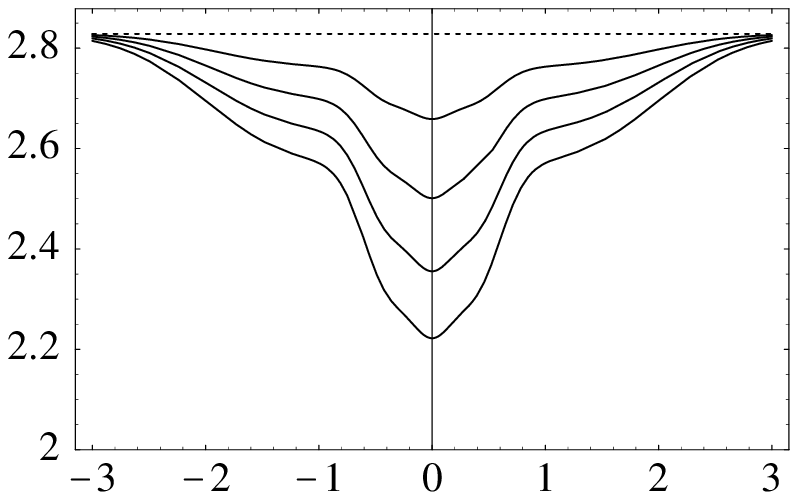}}}
\put(-8,0){${\cal B}^{\rm (max)}_{\eta}$}
\put(0.2,-4.5){$\cal J$}
\end{picture}
}
\vspace{2cm}
\caption{Plots of ${\cal B}^{\rm (max)}_{\eta}(\alpha,\beta,\alpha',\beta')$
in the case of the IPS state with $r=0.39$, $T=0.9999$, and $\varepsilon=1$
for different values of $\eta$; from top to bottom (solid lines): $\eta =
0.95, 0.9, 0.85$, and $0.8$. The dashed line corresponds to the value
$2\sqrt{2}$ obtained when $\eta = 1$. We set $\alpha = \beta = {\cal J}$ and
$\alpha' = \beta' = \sqrt{11}\,{\cal J}$, which maximize ${\cal B}$.}
\label{f:maxBparIPS}
\end{figure}

In the following we address the problem of evaluating the
maximum violation for the Bell states, the TWB and the IPS state when
a non-unit efficiency affects the displaced on/off photodetection.
First of all we note that
\begin{equation}
\Pi_{0,\eta}^2 = \sum_{k=0}^{\infty} (1-\eta)^{2k} \ket{k}\bra{k}
 = \Pi_{0, \eta(2-\eta)}\,,
\end{equation}
so that
\begin{align}
 {\cal E}_{\sf A}(\zeta) &= {\cal G}_{\eta}(\zeta) - {\cal
G}_{\eta(2-\eta)}(\zeta)\\
 {\cal E}_{\sf B}(\xi) &= {\cal Y}_{\eta}(\zeta) - {\cal
Y}_{\eta(2-\eta)}(\xi)\,.
\end{align}
In this way it is straightforward to evaluate ${\cal B}^{\rm
(max)}_{\eta}$ for the Bell states, the TWB and the IPS state. The results
are shown in the Figs.~\ref{f:B3D}--\ref{f:maxBparIPS}: in
Figs.~\ref{f:maxBpar}, \ref{f:maxBparTWB}, and \ref{f:maxBparIPS} we plot
${\cal B}^{\rm (max)}_{\eta}$ using the parametrization
$\alpha = \beta = {\cal J}$ and
$\alpha' = \beta' = \sqrt{11}\,{\cal J}$, which maximizes the Bell
parameter ${\cal B}_{\eta}$. It is worth noticing that for all the
considered states and for fixed ${\cal J}$ the limit
${\cal B}^{\rm (max)}_{\eta}$ is never reached;
on the other hand, even if the actual maximum violation, i.e., ${\cal
B}_{\eta, {\rm max}}$, is quite lower than the Cirel'son bound $2\sqrt{2}$,
it is relatively near to the {\em new} bound given by
Eq.~(\ref{BnewBound}).
\par
Notice that a similar analysis may be performed through the squared Bell
operator, which for $\Oop_{\sf k}^2(\zeta) \ne \iid$ is given by
\begin{multline}
\Bop^2 =
 \left( \Oop_{\sf A}^2( \alpha ) + \Oop_{\sf A}^2( \alpha' ) \right)
\otimes
\left( \Oop_{\sf B}^2( \beta ) + \Oop_{\sf B}^2( \beta' ) \right)\\
+ \left( \Oop_{\sf A}^2( \alpha ) - \Oop_{\sf A}^2( \alpha' ) \right)
\otimes
\left[ \Oop_{\sf B}( \beta ), \Oop_{\sf B}( \beta' ) \right]_{+}\\
- \left[ \Oop_{\sf A}( \alpha ), \Oop_{\sf A}( \alpha' ) \right]_{+}
\otimes
\left( \Oop_{\sf B}^2( \beta ) - \Oop_{\sf B}^2( \beta' ) \right)\\
+ \left[ \Oop_{\sf A}( \alpha ), \Oop_{\sf A}( \alpha' ) \right]
\otimes
\left[ \Oop_{\sf B}( \beta ) + \Oop_{\sf B}( \beta' ) \right]\,,
\label{general:B2}
\end{multline}
$[A,B]_{+}=AB+BA$ being the anti-commutator. As for Eq.~(\ref{BnewBound}),
the maximum value of Eq.~(\ref{general:B2}) depends on the state under
investigation and on the POVM itself.
\section{Concluding remarks}\label{s:remarks}
We have analyzed in details the nonlocality of several two-mode 
(entangled) states of  light by using a test based on 
displaced on/off photodetection. Nonlocality has been quantified 
through violation of CHSH inequality for the Bell's parameter. 
Effects due to non-unit quantum 
efficiency and nonzero dark counts have been taken into account. 
We found that unbalanced superpositions show larger nonlocality 
than balanced one when noise affects the photodetection process, 
and that twin-beam nonlocality is more robust than that of 
superpositions of few photon-number states. De-Gaussification 
by means of (inconclusive) photon subtraction is shown to 
enhance nonlocality of twin beams in the low energy regime. 
We have also shown that, since our measurement is described 
by a POVM rather than  a set of projectors, the maximum violation
the CHSH inequality cannot saturate the Cirel'son bound. A novel
state-dependent bound has been derived. 
\section*{Acknowledgments}
Fruitful discussions with A.~Ferraro are kindly acknowledged. This work 
has been partially supported by MIUR (FIRB RBAU014CLC-002).
\appendix
\section{Noisy on/off photodetection}
\label{a:onoff}
The action of an on/off detector in the ideal case is described by the 
two-value POVM $\{\Pi_0 = |0\rangle\langle 0|, \Pi_1 = {\mathbb I} -  
\Pi_0\}$, which represents a partition of the Hilbert space of the signal.
In the realistic case the performances of on/off photodetectors are
degraded by two effects. On one hand, one has non-unit quantum efficiency,
i.e., the loss of a portions of the incoming photons, and, on the other
hand, there is also  the presence of dark-count, 
i.e., by "clicks" that do not correspond to any incoming photon. In order 
to take into account both these effects we use a simple scheme described 
in the following.
\par A real photodetector is modeled as an ideal 
photodetector (unit quantum efficiency, no dark-count) preceded by a 
beam splitter (of transmissivity equal to the quantum efficiency $\eta$) 
whose second port is in an auxiliary excited state $ \nu$, which can 
be a thermal state, or a phase-averaged coherent state, depending on 
the kind of background noise (thermal or Poissonian). 
If the second port of the beam splitter is the vacuum 
$\nu = |0\rangle\langle 0|$ we have no dark-count; 
for the second port of the BS excited in a generic mixture $ \nu =
\sum_s \nu_{ss} |s\rangle\langle s|$ the POVM for the 
on/off photodetection is given by 
($\Pi_1 = {\mathbb{I}} -\Pi_0
$) 
\begin{eqnarray}
 \Pi_0 = \sum_{n=0}^\infty (1-\eta)^n \sum_{s=0}^\infty \nu_{ss} \: 
\eta^s \:\binom{n+s}{s}\: 
|n\rangle\langle n| \label{gendark}\;.
\end{eqnarray}
The density matrices of a thermal state and a phase-averaged coherent state 
(with $M$ mean photons) are given by 
\begin{eqnarray}
\nu_{\hbox{\tiny T}} &=& \frac{1}{M+1} \sum_s \left(\frac{M}{M+1}\right)^s 
\: |s\rangle\langle s| \\ \nu_{\hbox{\tiny P}} &=& e^{-M} \sum_s 
\frac{M^s}{s!}|s\rangle\langle s|  
\label{darkstate}\;. \end{eqnarray}
In order to reproduce a background noise with mean photon number $D$ 
we consider the state $\nu$ with average photon number $M=D/(1-\eta)$. 
\begin{widetext}
In this case we have 
\begin{eqnarray}
\Pi_{0,\eta,D}^{\hbox{\tiny T}} &=& \frac{1}{1+D} \sum_n \left( 1- 
\frac{\eta}{1+D}\right)^n
\:|n\rangle\langle n| \\  
\Pi_{0,\eta,D}^{\hbox{\tiny P}} &=& e^{-D} \sum_n \Bigg[(1-\eta)^n 
\: L_n \left(- D \frac{\eta}{1-\eta}\right)\Bigg] \:|n\rangle\langle n|
\label{darkpom}\;,
\end{eqnarray}
where ${\hbox{\tiny T}}$ and ${\hbox{\tiny P}}$ denotes thermal 
and Poissonian respectively, 
and $L_n(x)$ is the Laguerre polynomial of order $n$.
The corresponding Wigner functions are given by 
\begin{eqnarray}
\label{WPidark}
   W[\Pi^{\hbox{\tiny T}}_{0,\eta,D}](\alpha)
   &=& \frac{1}{\pi}\frac{2}{2(1+D)-\eta} \exp\left\{
 - \frac{2 \eta }{2(1+D)-\eta}\,|\alpha|^2   
   \right\}\,,\\
   W[\Pi^{\hbox{\tiny P}}_{0,\eta,D}](\alpha)
   &=& \frac{1}{\pi}\:\frac{2}{2-\eta} \exp\left\{
 - \frac{2 \eta}{2-\eta}(D+|\alpha |^2)  
   \right\}\: I_0 \left(\frac{4|\alpha |\sqrt{\eta D}}{2-\eta}\right)\;,
\end{eqnarray}
respectively, where $I_0(x)$ is the $0$-th modified Bessel function of
the first kind. For small $D$ the POVMs coincide up to first order, as
well as the corresponding Wigner functions.
\end{widetext}

\vfill

\begin{thebibliography}{99}
\bibitem{wer89} R.~F.~Werner, Phys.~Rev.~A {\bf 40}, 4277 (1989).
\bibitem{BW:PRL:99} K.~Banaszek, and K.~W\'odkiewicz, Phys. Rev. Lett. {\bf
82}, 2009 (1999).
\bibitem{FOP:2005} A.~Ferraro, S.~Olivares, and M.~G.~A.~Paris, {\em
Gaussian States in Quantum Information} (Bibliopolis, Napoli, 2005).
\bibitem{CHSH} J.~F.~Clauser, M.~A.~Horne, A.~Shimony, and R.~A.~Holt,
Phys. Rev. Lett. {\bf 23}, 880 (1969).
\bibitem{CH} J. F. Clauser, M. A. Horne, Phys. Rev. D \textbf{10}, 526 (1974)
\bibitem{bana:pra:02} K.~Banaszek, A.~Dragan, K.~W\'odkiewicz, and
C.~Radzewicz Phys. Rev. A {\bf 66}, 043803 (2002).
\bibitem{ips:PRA:67} S.~Olivares, M.~G.~A.~Paris, and R.~Bonifacio,
Phys. Rev. A {\bf 67},032314 (2003).
\bibitem{opatr:PRA:61} T.~Opatrn\'y, G.~Kurizki, and D.-G.~Welsch, Phys.
Rev. A {\bf 61}, 032302 (2000).
\bibitem{coch:PRA:65} P.~T.~Cochrane, T.~C.~Ralph, and G.~J.~Milburn, Phys.
Rev. A {\bf 65}, 062306 (2002).
\bibitem{nha:PRL:93} H.~Nha, and H.~J.~Carmichael, Phys. Rev. Lett. {\bf
93}, 020401 (2004).
\bibitem{garcia:PRL:93} R.~Garc\'\i a-Patr\'on S\'anchez et al., Phys.
Rev. Lett. {\bf 93}, 130409 (2004). 
\bibitem{ips:PRA:70} S.~Olivares, and M.~G.~A.~Paris, Phys. Rev. A {\bf 70},
032112 (2004).
\bibitem{ips:noise} S.~Olivares, and M.~G.~A.~Paris, quant-ph/0503104.
\bibitem{cirelsonB} B.~S.~Cirel'son, Lett. Math. Phys. {\bf 4}, 93 (1980).
\bibitem{landau} L.~J.~Landau, Phys. Lett A {\bf 120}, 54 (1987).
\bibitem{peres} A.~Peres, {\em Quantum Theory: Concepts and Methods}
(Kluwer, Dordrecht, 1993). 
\end{thebibliography}
\end{document}